\documentclass[12pt,preprint]{aastex}
\usepackage{epsfig,natbib,lscape,fullpage,indentfirst,color}
\shorttitle{Hot Jupiters}
\newcommand{\pdt}[1]{{{\partial #1}\over {\partial t}}}

\newcommand{\Mp}{M_p}

\begin{document}
\title{Atmospheric Dynamics of Short-period Extra Solar Gas Giant Planets I:}
\title{Dependence of Night-Side Temperature on Opacity} 
\author{Ian Dobbs-Dixon$^1$ \& D.N.C. Lin$^{1,2}$}
\affil{$^1$Department of Astronomy and Astrophysics,
University of California, Santa Cruz, CA 95064, USA}
\affil{$^2$Kavli Institute of Astronomy \& Astrophysics, Peking
University, Beijing, China}

\begin{abstract}
More than two dozen short-period Jupiter-mass gas giant planets have
been discovered around nearby solar-type stars in recent years,
several of which undergo transits, making them ideal for the detection
and characterization of their atmospheres. Here we adopt a
three-dimensional radiative hydrodynamical numerical scheme to
simulate atmospheric circulation on close-in gas giant planets. In
contrast to the conventional GCM and shallow water algorithms, this
method does not assume quasi hydrostatic equilibrium and it
approximates radiation transfer from optically thin to thick regions
with flux-limited diffusion.  In the first paper of this series, we
consider synchronously-spinning gas giants. We show that a full
three-dimensional treatment, coupled with rotationally modified flows
and an accurate treatment of radiation, yields a clear temperature
transition at the terminator. Based on a series of numerical
simulations with varying opacities, we show that the night-side
temperature is a strong indicator of the opacity of the planetary
atmosphere. Planetary atmospheres that maintain large, interstellar
opacities will exhibit large day-night temperature differences, while
planets with reduced atmospheric opacities due to extensive grain
growth and sedimentation will exhibit much more uniform temperatures
throughout their photosphere's. In addition to numerical results, we
present a four-zone analytic approximation to explain this dependence.
\end{abstract}
\maketitle

\section{Introduction}
Among the over 200 extra-solar planets discovered through a variety of
observational techniques including Doppler shifted spectral lines,
transit searches, and gravitational lensing, approximately $9\%$ have
orbital periods less then $4$ days. One of the more puzzling and
unexpected results to come from planet detections, these planets,
known as hot-Jupiter's or Pegasus planets, receive approximately
$10^4$ times more energy from stellar irradiation than from internal
heating. This energy input greatly modifies the thermal structure and
dynamics of the atmosphere. Given the short orbital period of these
planets, tidal forces also play a strong role, quickly circularizing
their orbits and synchronizing their spin and orbital frequencies
\citep{goldreich1966, dobbsdixon2004}. Consequently, one side of the
planet perpetually faces the host star, reaching equilibrium
temperatures of over $1000 \mathrm{ K}$, while night-side temperatures
are determined by the ability of the planet to redistribute energy. In
the absence of heat redistribution, the night-side of the planet would
remain at $\approx 100\mathrm{ K}$ after a few Gyr of cooling and
Kelvin-Helmholtz contraction. For comparison, Jupiter has a rotational
period of $11$ hours and its internal luminosity is approximately
equal to the energy it receives from the Sun. Its photospheric
temperature is quite uniform in both longitude and latitude.

The small semi-major axis of hot-Jupiter's provide several new methods
for exploring their structural parameters and atmospheres including
primary transits \citep{charbonneau2007}, secondary eclipses
\citep{deming2005, charbonneau2005, deming2006}, and spectroscopy
\citep{charbonneau2002, brown2002, deming2005_2, grillmair2007,
richardson2007, swain2007}.  Detection of a primary eclipse, as the
planet passes between the star and the Earth, yields measurements of
planetary radii and masses that can be used to infer structural
parameters of the planets. The secondary eclipse, caused by the
passage of the planet behind its host star, provides a direct
measurement of the day-side temperature of the objects by measuring
the decrement in infrared flux. As of the writing of this paper, $14$
planets have been detected via the primary transit method, while $3$
of those have also been detected via their secondary eclipse. The
day-side temperatures of the objects found from secondary eclipse
measurements are found to be $T = 1130\pm150 \mathrm{K}$ for HD209458b
\citep{deming2005}, $T = 1060\pm50 \mathrm{K}$ for TrES-1
\citep{charbonneau2005}, and $T = 1117\pm42 \mathrm{K}$ for HD189733b
\citep{deming2006}. Measurements of HD209458b and HD189733b were done
using the Spitzer Space Telescopes $24\mu m$ and $16\mu m$ band-passes
respectively. The estimate of temperature is then sensitive to both
the ratio of stellar and planetary radii and stellar temperature.
TrES-1 was observed in both the $4.5\mu m$ and $8\mu m$ bands, and
temperatures were derived by assuming the planet emits as a
blackbody. In addition, recent non-transiting observations of
$\nu-$Andromeda b with the Spitzer Space Telescope, indicate a
substantial orbital phase dependence in the flux at $24\mu m$
\citep{harrington2006}.

A number of groups have used one-dimensional, plane-parallel radiative
transfer models to calculate the emergent spectra of these
hot-Jupiter's \citep{fortney2005, burrows2005, seager2005, barman2005,
fortney2006_2, burrows2006_2, fortney2006}. These approaches
simultaneously solve the equations of radiative transfer, radiative
equilibrium, and hydrostatic balance to determine the pressure,
temperature, and spectral energy distribution as a function of radial
position in the atmosphere. The input parameters include the
temperatures at the bottom of the atmosphere and the incident stellar
energy spectrum at the top. The temperature at the bottom is
determined primarily by the internal luminosity, and ideally it should
be taken from one-dimensional evolutionary and structural
models. However, given the dominate influence of the incident energy
near the surface, the choice of the bottom does not noticeably
influence the emergent spectra.

In hot-Jupiter's, the intense irradiation of the day-side drives
strong thermal winds toward the night-side. The resulting temperature
in the upper atmosphere depends on the ability of these winds to
redistribute the stellar irradiation, and in general should be a
function of both longitude and latitude. In the absence of dynamical
models, this redistribution of incident energy at the upper layers is
usually set to be some fraction of the incident irradiation, to
represent the degree of energy re-distribution. The value of this
parameter is a major uncertainty in these radiative models, and
authors have computed a number of cases with varying degrees of
re-distribution. In addition, such parameterization neglects the role
of advection and radiative transfer within lower levels of the
atmosphere, which are also important in determining the final
pressure-temperature profiles.

In an attempt to address the dynamical redistribution of incident
stellar energy within the atmosphere, there have been a number of
dynamical models of the atmospheres of hot-Jupiter's. These models
utilize a variety of methods employing various simplifications,
including solving the primitive equations
\citep{showman2002,cooper2005}, the shallow water equations
\citep{cho2003,menou2003,langton2007}, the equivalent-barotropic
equations \citep{cho2006}, and two-dimensional hydrodynamical
equations \citep{burkert2005}. These models predict a wide range of
behavior. More details on these models are presented in
\S\ref{sec:comparison}, in comparison with the model presented in this
paper. One group \citep{fortney2006} has attempted to couple the
spectral and dynamical models by utilizing pressure and temperature
profiles derived directly from the simulations of
\citet{cooper2005}. Unlike previous models, these atmospheres are not
iterated to achieve radiative-convective equilibrium.

In this paper, we present the results obtained from a three-
dimensional, hydrodynamic simulation based on flux-limited radiative
transfer models of hot-Jupiter's for a variety of rotation rates and
opacities. In \S\ref{sec:method}, we present the basic equations,
numerical methods, and initial conditions. In \S\ref{sec:results} we
present our results and analysis for both rotating and non-rotating
flows. We also study the effects of changing opacity on the dynamics
and heat distribution. In \S\ref{sec:comparison} we include a detailed
comparison with previous dynamical models in an attempt to highlight
the consequences of making certain simplifying assumptions. We conclude
in \S\ref{sec:discussion} with a discussion.

\section{Numerical Method}
\label{sec:method}

\subsection{Flux-Limited Radiative Hydrodynamical Model}
\label{section:numerical}
We cast our numerical model in spherical coordinates
$\left(r,\phi,\theta\right)$, where $\phi$ is the azimuthal angle (or
longitude) and $\theta$ is the meridional angle (or latitude) measured
from the equator. Including both the Coriolis ($2{\bf \Omega\times
u}$) and centrifugal forces (${\bf\Omega\times} \left({\bf\Omega\times
r} \right)$), the equations of motion for the fluid can be written as
\begin{equation}
\pdt{\rho} + \nabla\cdot\left(\rho{\bf u}\right) = 0
\label{eq:continuity}
\end{equation}

\begin{equation}
\pdt{{\bf u}} + \left({\bf u}\cdot\nabla\right) {\bf u}= -
\frac{1}{\rho}\nabla{P} + \frac{1}{\rho}\nabla{\Phi} -2{\bf
\Omega\times u} - {\bf\Omega\times}\left({\bf\Omega\times r}\right)
\label{eq:momentum}
\end{equation}

The rotation frequency is given by $\Omega$, and the gravitational
potential, $\Phi=-{G\Mp\over r}$, varies only in the radial
direction. We neglect explicit viscosity, but some degree of numerical
viscosity is inevitable. We also include the curvature terms in
$\left({\bf u}\cdot\nabla\right) {\bf u}$. We solve equations
(\ref{eq:continuity}) and (\ref{eq:momentum}) on a stagged grid, where
scalars are defined in the center of cells and vectors on cell
boundaries. This method yields second-order spatial accuracy. Given
the decreasing grid size near the pole, our computational domain is
limited to $\theta =\pm 70^{\circ}$. Although excluding this region
neglects an avenue for energy re-distribution, even a modest amount of
rotation (approximately $2$ to $4\mathrm{ days}$ for the
hot-Jupiter's) will cause the dominate flow patterns to be
concentrated near the equator. We simulate the entire azimuth of the
planet, instituting periodic boundary conditions at $\phi = 0$ and
$2\pi$.  The radial extent of the domain extends from $7.95\times10^9
\mathrm{ cm}$ to $8.65\times 10^9 \mathrm{ cm}$, corresponding to
$1.06$ to $1.2 R_J$.  The pressure scale height on the day-side is
approximately $340$ km.

Our numerical radiative transfer scheme is capable of following the
temperature and radiation energy independently, linking them with a
given heating/cooling function. However, this degree of sophistication
is not necessary for calculations in which the gas is close to thermal
equilibrium. Instead, we use a one-fluid approximation where radiation
energy density is a simple function of temperature, $E = a T^4$. For
the temperatures considered here, the radiative energy density is much
lower than the thermal energy density. The internal and radiation
energy equations reduce to
\begin {equation}
\left[ \pdt{\epsilon} + ({\bf u}\cdot\nabla) \epsilon \right] = - P
\, \nabla \cdot {\bf u} - \nabla \cdot {\bf F}
\label{eq:energy}
\end{equation}
where $\epsilon=c_{v} \rho T$ is the internal energy density, $T$ is
the temperature, $c_v$ is the specific heat, and ${\bf F}$ is the
radiative flux.

To calculate the radiative flux we use the flux limited radiation
transfer approximation of \citet{levermore1981},
\begin {equation}
{\bf F} = - \lambda {\frac{c}{\rho\kappa}} \nabla E,
\end {equation}
where $\lambda$ is a non-constant flux limiter which prescribes the
relationship between the radiative flux and the radiative energy
gradient. We use the flux limiter developed by \citet{levermore1981},
given by
\begin{equation}
\lambda = \frac{2+R}{6+3R+R^2},
\label{lambda}
\end {equation}
where
\begin {equation}
R = \frac{1}{\rho \kappa} \frac{| \nabla E |}{E}
\label{Rdef}
\end {equation}
compares the scale height of the radiation energy to its mean free
path. This overall procedure allows for an accurate treatment of both
optically thick and thin regions; in the optically thick {\it
diffusion} limit $R \rightarrow 0$, $\lambda \rightarrow 1/3$, and
${\bf F} \rightarrow - \, c \nabla E / 3 \rho \kappa $, while in the
optically thin {\it streaming} limit $R \rightarrow \infty$, $\lambda
\rightarrow 1/R$, and ${\bf F} \rightarrow c E$. While the flux
limiter $\lambda$ is an approximation, it is quite accurate in both
the optically thick and thin limits.

In order to follow the evolution on a long time-scales, the radiative
portion of the energy equation is solved {\it implicitly}. For
quasi-static radiative conditions such as those considered here, the
equations can be advanced much more rapidly than if they were
restricted by a radiative time-step. We use the successive
over-relaxation method (SOR) to solve the $\nabla \cdot {\bf F}$
portion of equation (\ref{eq:energy}), alternately updating even and
odd grid cells. The limiting factor for the numerical time-step then
becomes the Courant condition.

The advection scheme, described in \citet{hawley1984} and
\citet{kley1987}, is an extension of the first-order van Leer
\citep{vanleer1977} method known as the 'mono-scheme'. It employs
operator-splitting, where the finite difference equations are split
into parts, which are then solved separately always using the latest
values of the variables. The scheme yields semi-second order temporal
accuracy and allows for the accurate resolution of discontinuities in
the fluid flow with limited diffusion.

Finally, we use an ideal gas equation of state for the pressure,
$p=({\sf R}_{G}\rho T)/\mu$, with specific heat $c_{v}={\sf R}_{G} /
(\mu (\gamma-1))$ where ${\sf R}_{G}$ is the gas constant, $\gamma
=\frac {7}{5}$, and the mean molecular weight is fixed at $\mu =
2.3$. Although the temperatures in the hydrodynamic models do become
hot enough to dissociate hydrogen molecules in some locations within
the planet, the region which we are most concerned with is well
described by a constant molecular weight. Radiative opacities are
found using the tables of \citet{pollack1985} for lower temperatures
coupled with \citet{alexander1994} for higher temperatures. These are
Rosseland mean opacities and include the effects of atomic, molecular,
and solid particulate absorbers and scatters. The opacity is one of
the largest uncertainties. The effect of composition, clouds,
settling, wavelength dependence, grains' condensation, sublimation, 
collisional growth, and sedimentation are but a few of the parameters
that alter the magnitude of the opacity. To address this uncertainty,
we explore the effect of varying opacity in \S\ref{sec:opacity}.

\begin{figure}
\plotone{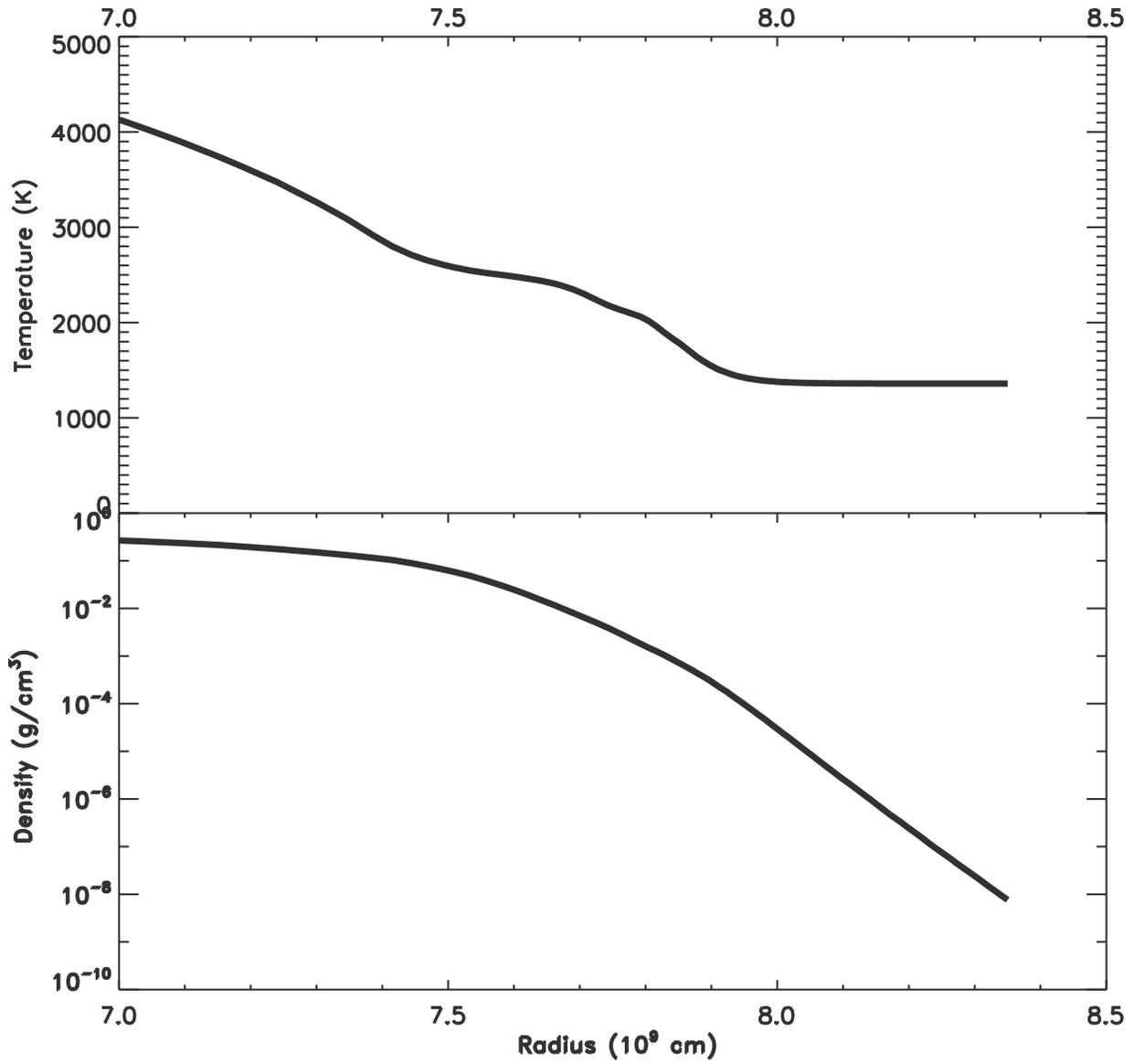}
\caption{Spherically symmetric temperature and density profiles from
the one-dimensional evolutionary models of \citet{bodenheimer2001}
used as an initial condition for the simulations presented here. The
models were run for $4.5 \mathrm{Gyr}$, during which time the
temperature at the upper boundary was held at $1200\mathrm{K}$ to
simulate the irradiation from the central star.}
\label{fig:initmod}
\end{figure}

Our initial density and temperature profiles are taken from a
one-dimensional planetary evolutionary code. For details on the model,
see model B1 of \citet{bodenheimer2001}. Use of this model implies
that the simulations are initially spherically symmetric. Although
unrealistic, we allow simulations to run for sufficient time to relax
into equilibrium configurations; the details of the initial state are
lost. Figure (\ref{fig:initmod}) shows the initial temperature and
density profiles from the one-dimensional model near the top of the
planet. The model follows a $0.63M_J$ planet for $4.5
\mathrm{Gyr}$. The upper boundary was held at a fixed temperature of
$1200\mathrm{K}$ to simulate the effects of irradiation from the
central star. As noted in \citet{cho2006}, a quiescent initial start
neglects the effects of pre-established small-scale structures such as
eddies and jets. Given that we lack the resources to simultaneously
study these small scale effects and full three dimensional effects, we
assume that the dynamics will be dominated by the large scale
anisotropic heating imposed by the stellar irradiation.  In contrast
to the shallow-water approximation, specific vorticity can be
generated in our full 3D radiative hydrodynamic simulations.
Therefore, baroclinic instabilities can be excited spontaneously and
can lead to the generation of structure down to the resolution length
scales.

To represent the impinging radiation, we set the temperature at the
upper boundary to $T = \mathrm{max}\left[1200\mathrm{K}
\left(cos\left(\phi\right)cos\left(\theta\right)\right)^{1/4},
100\mathrm{K}\right]$,
approximating the stellar radiation field with a maximum at the
sub-stellar point, and a night-side held at $100\mathrm{ K}$. The
night-side temperature is chosen to be consistent with the
photospheric temperature derived in non-irradiated planetary evolution
models. This imposed temperature distribution will increase the scale
height of the atmosphere on the day-side, while cooling and
contracting it on the night-side. An interesting result is a planet
that is no longer spherical; the scale-height on the day-side is
somewhat larger than on the night-side. Given the computational
difficulties of modeling the extremely low density regions in the
upper atmosphere, we impose a movable upper boundary at the location
where $\rho<10^{-9}\mathrm{g/cm^3}$. At the bottom boundary we specify
the temperature flux taken from our initial one-dimensional models. As
mentioned above, the energy input from the star overwhelms the planets
intrinsic luminosity in the upper atmosphere, so the choice of this
boundary condition is not critical. We have run several simulations
with fixed temperature to verify the validity of this approximation,
and found little difference in the resulting energy distribution or
dynamics.

\section{Results of Numerical Simulations}
\label{sec:results}

\subsection{The Non-Rotating Model}
\begin{figure}
\plotone{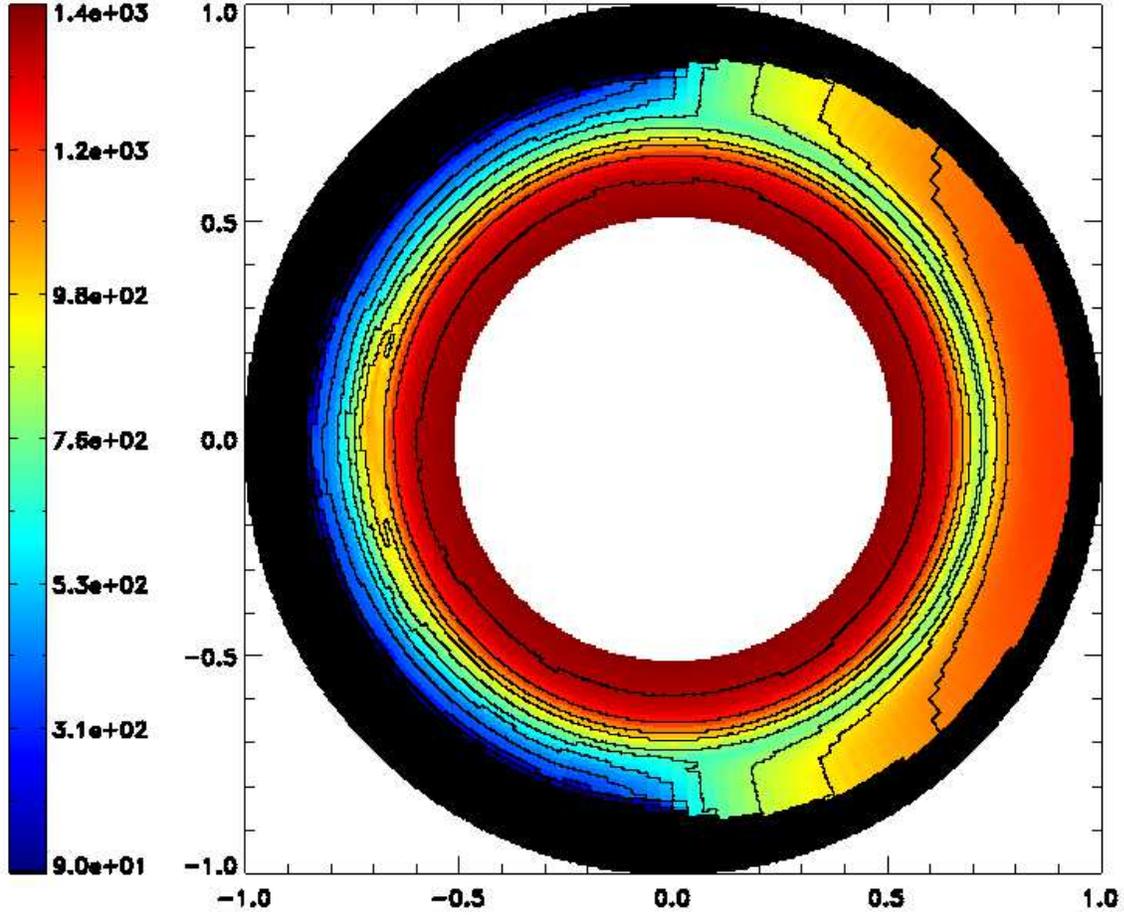}
\caption{The temperature distribution at the equator of a non-rotating
hot-Jupiter. The inner boundary is assumed to be spherically
symmetric, with an outward energy flux fixed from the initial
one-dimensional model. The outer black area represents regions with
$\rho < 10^{-9} g/cm^3$, outside our movable boundary.}
\label{fig:norot_teq}
\end{figure}

As a reference, we first present the results from an idealized and
artificial non-rotating model in which the gas giant is 
subject to one-sided stellar irradiation. The azimuthal pressure
gradient associated with the imposed temperature contrast drives
strong winds toward the night-side of the planet. The flow pattern is
symmetric with respect to the planet-star line, and fluid flows to the
night-side around both sides of the planet. As the material moves to
the night-side, it achieves maximum velocities of $\sim 3 \mathrm{
km/s}$ near the terminators ($\phi=\frac{\pi}{2}$ and
$\phi=\frac{3\pi}{2}$). This velocity corresponds to Mach numbers of
up to $2.7$. A substantial portion of energy is released in these
shocked regions. By the time the two symmetric flows converge on the
night-side they have both undergone substantial cooling and sink
radially inward, initiating a return flow at depth. In Figure
(\ref{fig:norot_teq}) we show the temperature distribution in the
equatorial plane. Despite the high winds, a clear day/night difference
persists. Also evident is the cool region at depth, which is due to
the confluence of the negative temperature gradient from the interior,
the positive temperature gradient near the surface, and the cool flow
returning to the day-side, completing the two, approximately
symmetric, azimuthal convection cells.

\begin{figure}
\begin{center}
\includegraphics[height=8.0cm,width=12.0cm]{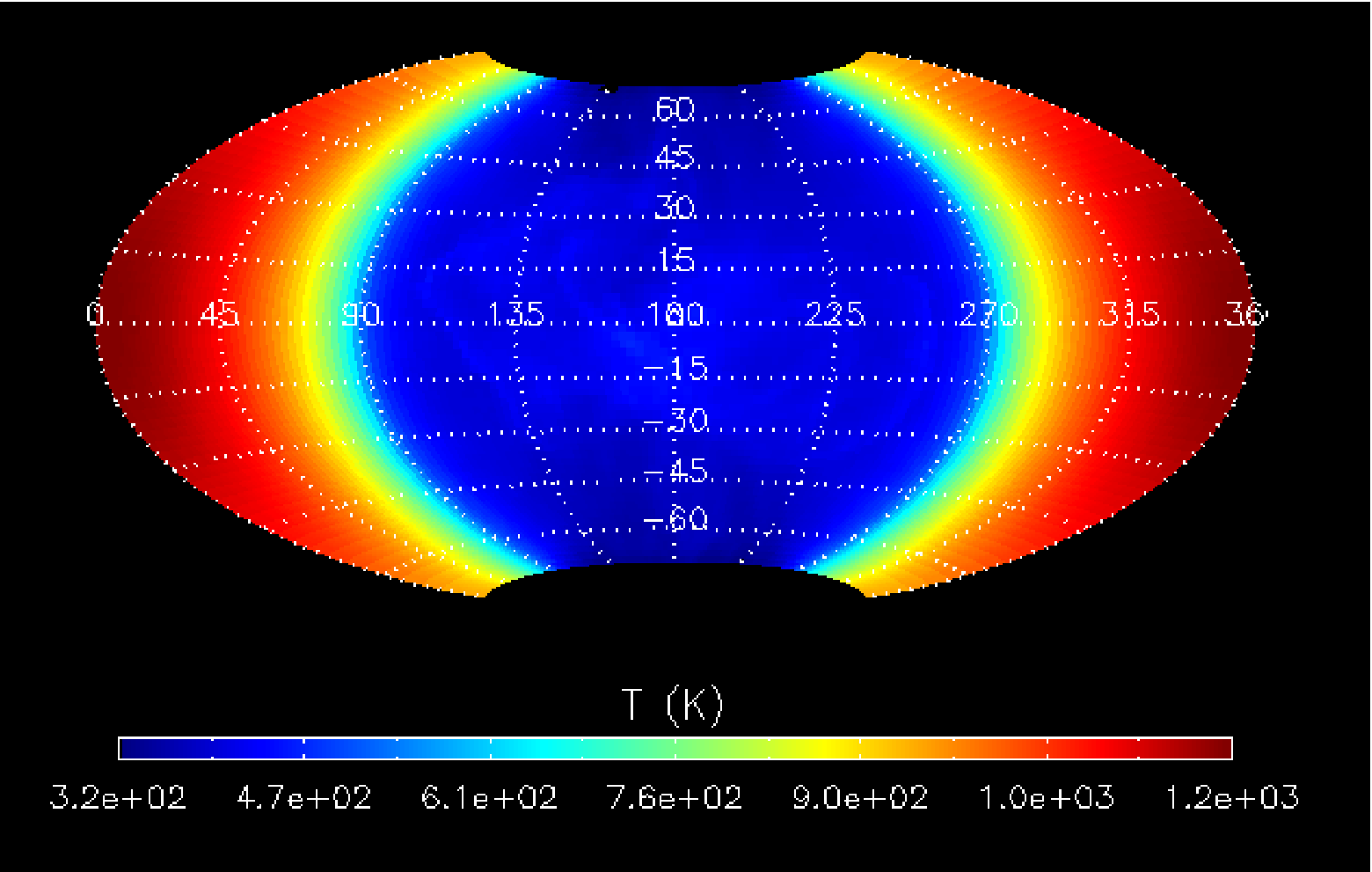}
\includegraphics[height=8.0cm,width=12.0cm]{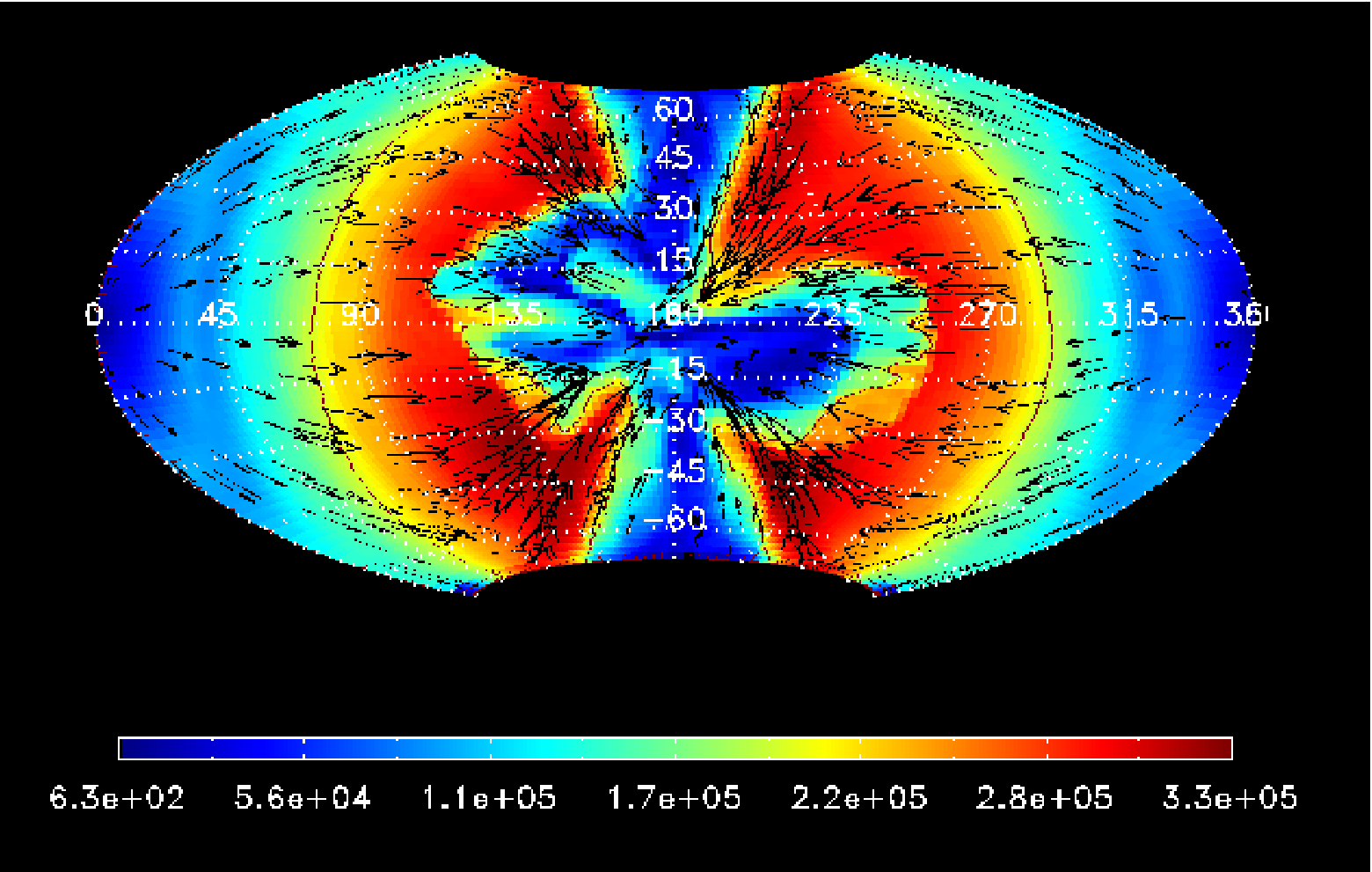}
\end{center}
\caption{The temperature (upper panel) and velocity (lower panel) at
the photosphere of a non-rotating planet. A strong temperature
gradient between the day-side ($\phi = 0^\circ$) and the night-side
($\phi=180^\circ$) is evident. The regions of largest $\nabla T$
correspond to fluid motions of $> 3 km/s$.}
\label{fig:norot_tnvphoto}
\end{figure}

Figure (\ref{fig:norot_tnvphoto}) shows the temperature and velocity
at the photosphere ($\tau=\frac{2}{3}$) of the planet. Given the
changing scale height near the surface of the planet, these plots do
not represent a constant radial surface. Moving from the day to
night-side, a clear increase in the velocity near the terminator is
evident, followed by a drop at the convergent point. It should also be
noted that the convergence point is quite dynamical, oscillating in
both longitude and latitude. However, despite these oscillations, two
distinct convective cells remain. Night-side temperatures range from
$300-550 \mathrm{K}$, with an average night-side temperature
($\frac{\pi}{2}<\phi<\frac{3\pi}{2}$) of $414\mathrm{K}$. Also evident
in Figures (\ref{fig:norot_teq}) and (\ref{fig:norot_tnvphoto}) is an
area of increased temperature near $180^{\circ}$. As the two
converging flows meet on the night-side, compressional heating drives
up the temperatures, and thus the local scale height. The resulting
increase is small, but can be seen both at depth and near the
photosphere.

\subsection{A Planetary Model with 3-Day Rotation}
It is widely believed that tidal forces within the atmospheres and the
envelope of hot-Jupiter's drive them to tidally locked spin
configurations on time-scales much shorter than the main sequence life
span of their host stars.  In this synchronous state, the planets'
spin frequency equals their orbital frequency. With this assumption,
hot-Jupiter's spin with periods on the order of $3\mathrm{-days}$. In
comparison to our own giant planets, this is a relatively slow spin
rate. Nevertheless, the associated Coriolis force significantly alters
the resulting flow dynamics and may have implications concerning the
ability of the planet to fully synchronize its spin. Another crucial
consideration that came to light during this study was the effects
associated with the initialization of the rotation. The
one-dimensional initial models described in Section
(\ref{section:numerical}) were non-rotating models, and the rate that
we chose to turn on the rotation had observable effects. We will
explore this effect in a subsequent paper.

\subsubsection{The Day-Side Isothermal Surface}
\begin{figure}
\plotone{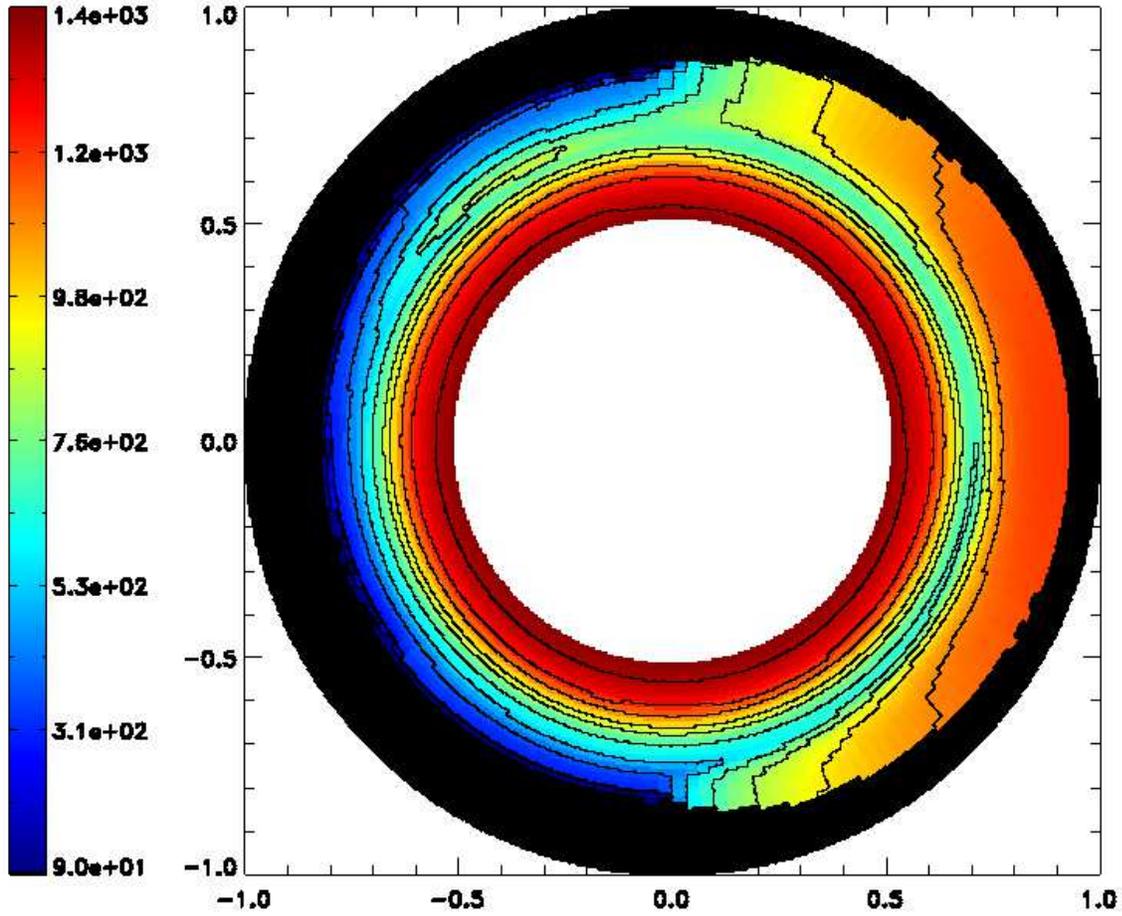}
\caption{The temperature distribution at the equator of a planet
rotating with a period of $3 \mathrm{ days}$. The day-side is
characterized by a large isothermal area near the top, the extent of
which falls off with increasing longitude. A cool region at depth is
also evident due to the combined effects of a negative temperature
gradient from the internal heating, a positive temperature gradient
from the irradiation, and a cooling return flow from the night-side.}
\label{fig:3d_teq}
\end{figure}

In Figure (\ref{fig:3d_teq}) we show the temperature distribution in
the equatorial plane for a simulation rotating with a period of $3
\mathrm{-days}$. The sub-solar point on the day-side is characterized
by an radially-extended nearly isothermal region with an effective
day-side temperature $T_d \sim 1200K$.  Upon adjusting to a
hydrostatic equilibrium, a slightly negative temperature gradient is
established so that the reprocessed stellar radiation can penetrate
into the planetary envelope.  Nevertheless the day-side photosphere is
essentially isothermal with a density profile
\begin{equation}
\rho (r) = \rho (r_b) {\rm exp}\left[ - (g \mu / {\sf R}_{G} T_d)
(r-r_b)\right],
\end{equation}
where $r_b$ is a planet's radius at the base of the isothermal region
and $g = G M_p/ r_b^2$ is the surface gravity of the planet.  The
imposed stellar heating falls off as a function of longitude, thus the
radial extent of this isothermal region decreases with the inclination
angle between the local zenith and the position of the host star
overhead.

\subsubsection{Azimuthal Effective-Temperature Distribution}
\begin{figure}
\begin{center}
\includegraphics[height=8.0cm,width=12.0cm]{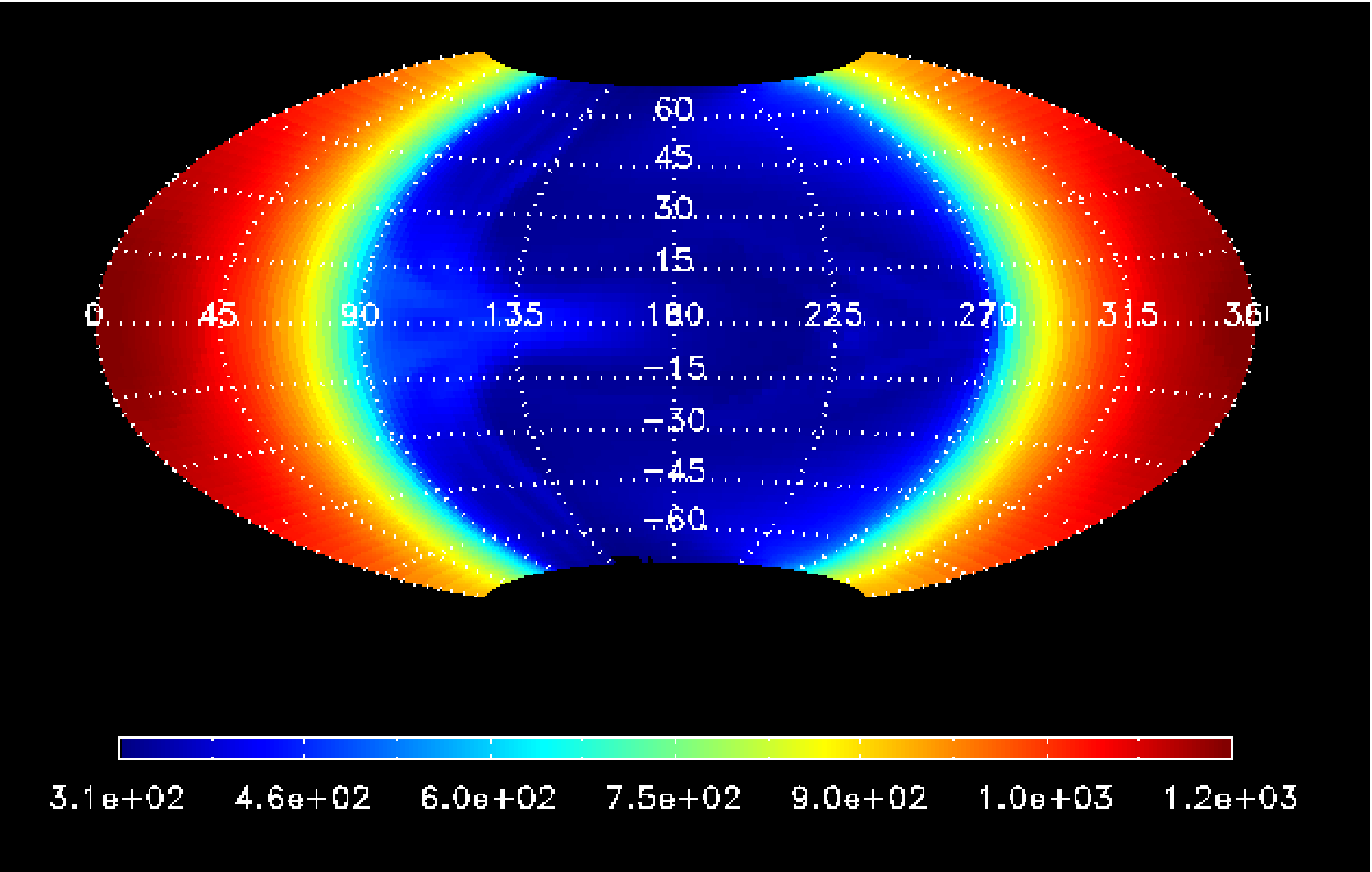}
\includegraphics[height=8.0cm,width=12.0cm]{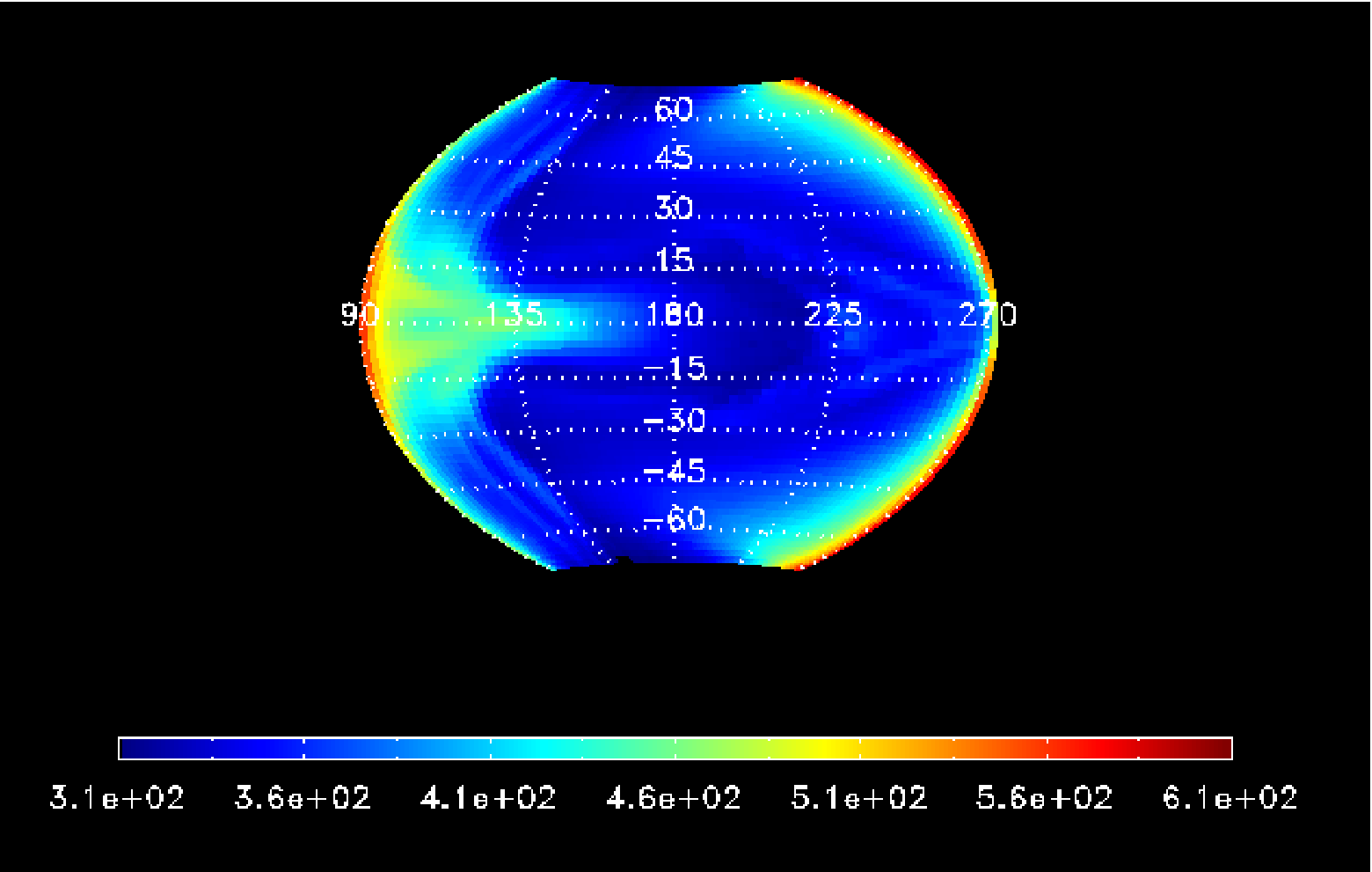}
\end{center}
\caption{The temperature at the photosphere of a planet rotating with
a period of $3\mathrm{ days}$. The upper panel shows the distribution
over the entire planet, while the lower panel highlights the
temperature structure on the night-side from $\phi=\frac{\pi}{2}$ to
$\phi=\frac{3\pi}{2}$. A clear day-night delineation persists, despite
complicated dynamical structure, due to substantial radiation near the
terminators.}
\label{fig:3d_tph}
\end{figure}

Below the isothermal photosphere, a cooler region at lower depths is also
evident on the day-side, associated in part with a cool return flow
from the night-side. Figure (\ref{fig:3d_tph}) shows temperature
distribution both throughout the entire photosphere, and focusing on
structure on the night-side. Despite the added effect of rotation, a
clear day-night delineation is still apparent, with the night-side
characterized by effective temperatures $T_n$ from $310$ to
$500\mathrm{ K}$. The average night-side temperature is $380\mathrm{
K}$, slightly smaller then the non-rotating simulation with the same
opacities. This slight decrease in average temperature is due to
increased cooling associated with rotationally modified flows
discussed in the next sub-section. Slightly hotter regions near the
terminators, associated with jets from the day-side, are apparent with
temperatures reaching $\sim 500\mathrm{ K}$.

\subsubsection{Thermal Current and the Corilois Effect}
\begin{figure}
\begin{center}
\includegraphics[height=8.0cm,width=12.0cm]{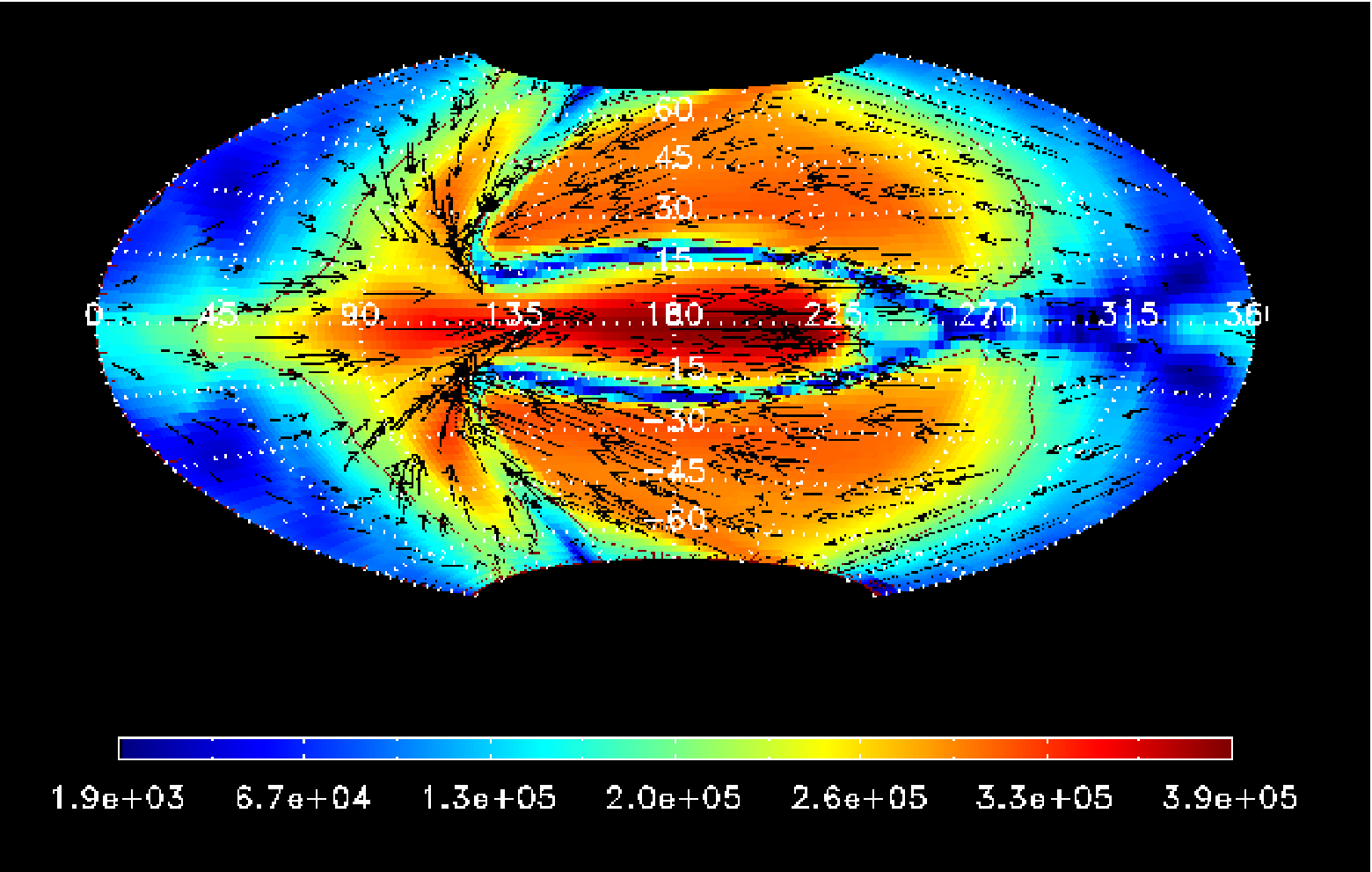}
\includegraphics[height=8.0cm,width=12.0cm]{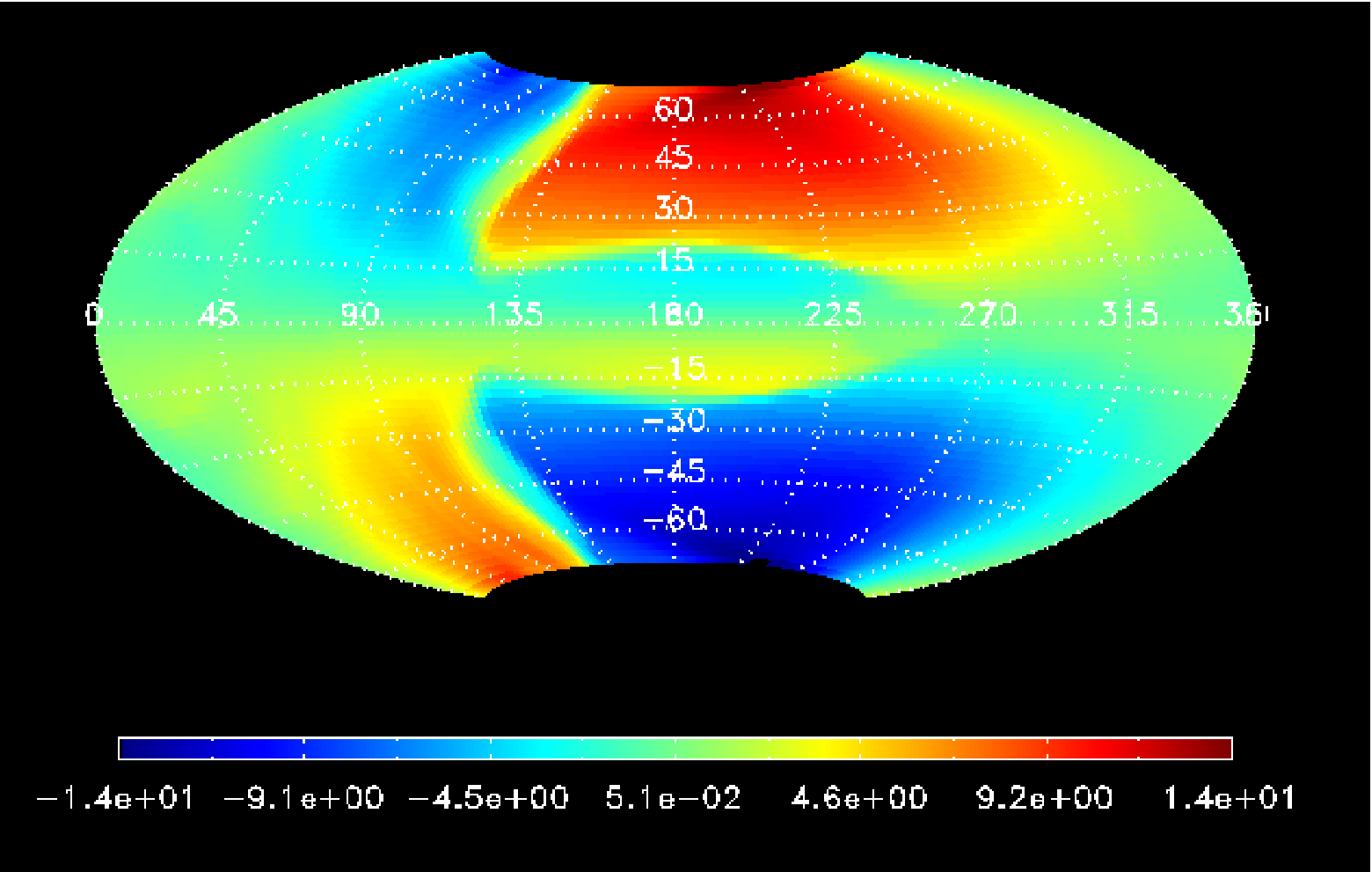}
\end{center}
\caption{The upper panel shows the magnitude of the velocity at the
photosphere, given by $\left|v\right| =
\sqrt{v^2_\phi+v^2_\theta}$. The lower panel shows the latitudinal
component of the Coriolis effect ($-2\Omega
v_{\phi}sin\left(\theta\right)$) at the photosphere of a planet
spinning at $3\mathrm{ days}$. The direction of the latitudinal
Coriolis force is different for eastward and westward moving material,
causing eastward flows to be focused toward the equator, while
westward flow is funneled toward the poles.}
\label{fig:v_coriolis_ph}
\end{figure}

The upper panel of Figure (\ref{fig:v_coriolis_ph}) shows the velocity
magnitude at the photosphere ($\left|v\right| = \sqrt{v^2_\phi+
v^2_\theta}$). As in the non-rotating model, material is moving quite
rapidly, reaching speeds of $\sim 4\mathrm{ km/s}$ near the
terminators.  Eastward (prograde or $v_\phi >0$, {\it i.e.} in the
same direction as the unperturbed spin) moving material appears to be
funneled from the sub-tropical latitudes ($\vert \theta \vert > 0$)
into an equatorial jet near $\phi=\frac{\pi}{2}$, while westward
(retrograde or $v_\phi <0$) moving material is pushed from the
subtropical zone toward the poles near $\phi=\frac{3\pi}{2}$. It is
these flow structures, rapidly advecting energy from the day-side,
that account for the hotter regions seen on the night-side in Figure
(\ref{fig:3d_teq}). To understand this structure, we must evaluate the
${\bf\hat{\theta}}$ component of the Coriolis force, given by
$-2\Omega v_{\phi}sin\left(\theta\right)$, which is shown in the lower
panel of Figure (\ref{fig:v_coriolis_ph}). Although material near the
equator doesn't feel any Coriolis force in this direction, it is clear
that material at higher and lower (sub-tropical) latitudes does. The
asymmetry imposed by the rotation ({\it i.e.} the azimuthal
velocities) causes the fluid moving in eastward and westward
directions to behave significantly different then in the non-rotating
case.

The approximate magnitude of velocity can be estimated from equation
(\ref{eq:momentum}), considering only the pressure gradient
term. Assuming an approximately constant acceleration around to the
night-side of the planet given by $a \simeq -\frac{1}{\rho}\nabla P$,
the velocity at the terminator should be given by
\begin{equation}
v_T \simeq \left[\frac{2 \gamma {\sf R}_{G}}{\mu}
\left(T_d-T_n\right)\right]^{1/2}.
\label{eq:vel_terminator}
\end{equation}
Given a day-side temperature of $T_d \simeq 1200K$ and an average
night-side temperature of $T_n \simeq 350K$, flows should achieve Mach
numbers of $\sim 2$ near the terminator. The sound-speed 
$( \gamma {\sf R}_{G}
T/\mu )^{1/2} \sim 1.7$ km/s at the terminator, yielding a local Mach
number, as predicted, of $\sim 2$.

\begin{figure}
\begin{center}
\includegraphics[height=8.0cm,width=9.0cm]{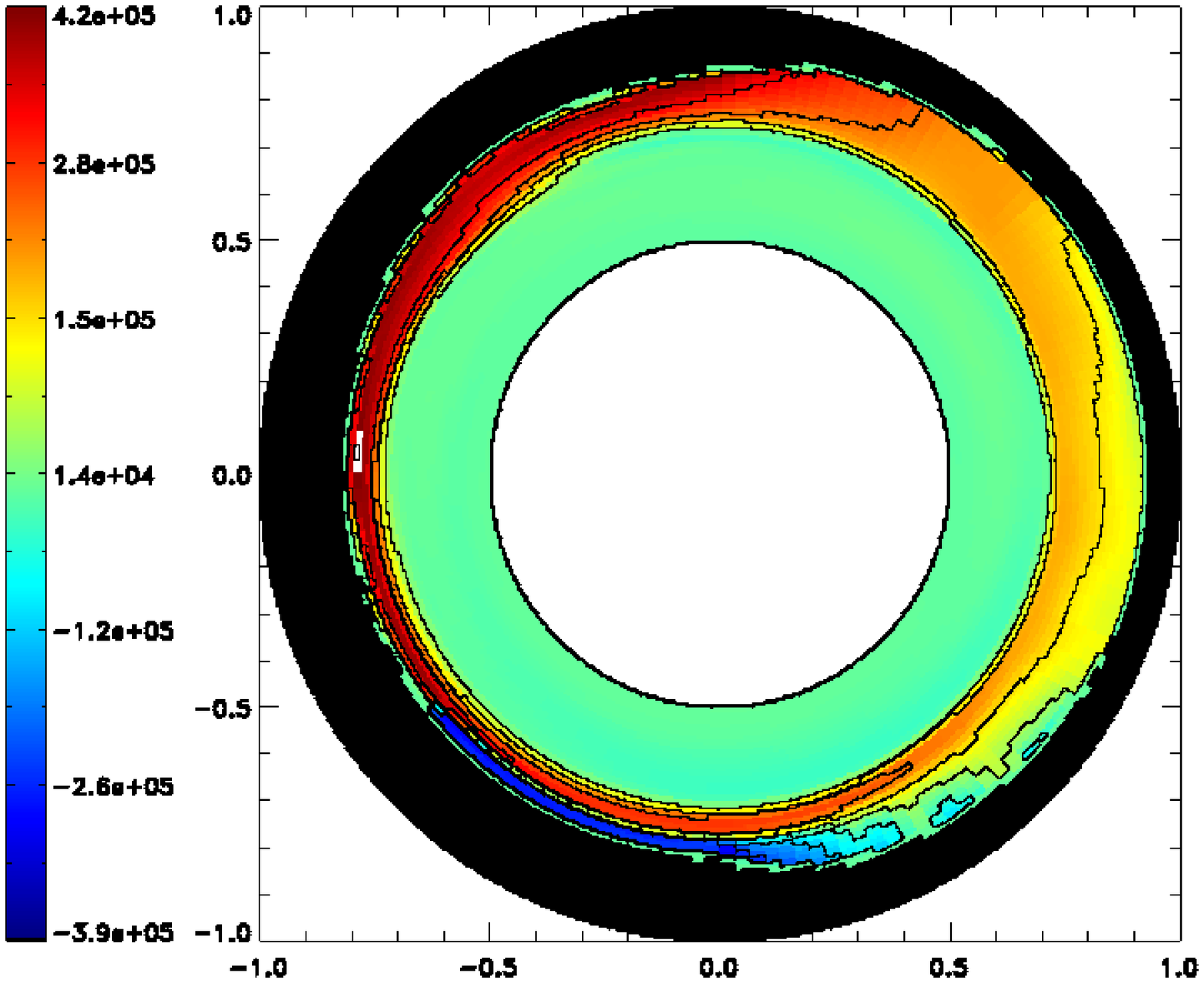}
\includegraphics[height=8.0cm,width=9.0cm]{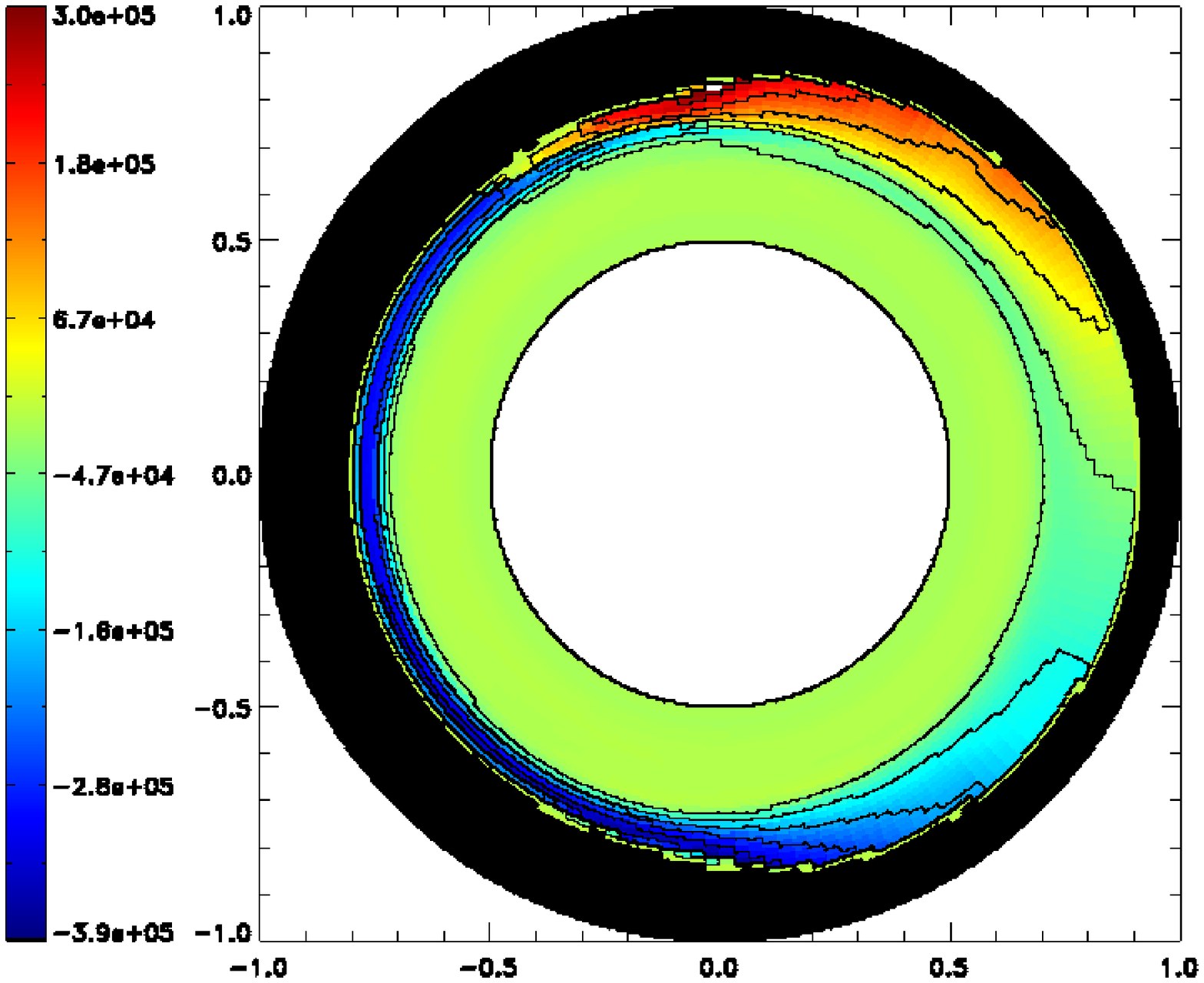}
\end{center}
\caption{The velocity distribution at the equator (upper panel) and a
latitude of $\theta=35^{\circ}$ (lower panel) of a planet rotating
with a period of $3 \mathrm{ days}$. At the equator, eastward flow
(red) is able to circumnavigate the planet at depth, while at higher
and lower latitudes, it is westward flow (blue) that is able to
traverse all the way around the planet. The sinking of one flow under
the other is due to different degrees of cooling due to rotationally
altered dynamics.}
\label{fig:3d_veq_high}
\end{figure}

Also evident in the plot of velocity magnitude in Figure
(\ref{fig:v_coriolis_ph}) is the marked decrease in velocity where the
eastward and westward flows converge. Neither flow is able to
instigate circumplanetary flow at the surface. Figure
(\ref{fig:3d_veq_high}) shows an equatorial slice of the azimuthal
velocity $v_{\phi}$ at the equator (upper panel) and at higher
latitudes (lower panel). It is evident from this plot that the
eastward moving flow does continue around the planet at depth near the
equator, while the westward moving fluid continues around the planet
at higher (and lower) latitudes. Because of the effects of rotation
shown in Figure (\ref{fig:v_coriolis_ph}), the convergence point is
near $\phi=\frac{5\pi}{4}$ for the equatorial flow, and near
$\phi=\frac{3\pi}{4}$ for flows at higher and lower latitude. This
flow pattern implies that, upon converging, one of the two flows has
undergone substantially more cooling then its counterpart. Thus, at
the equator, when the eastward flow encounters the westward flow near
$\phi=\frac{5\pi}{4}$ (past ``mid night''), the former is cooler and
sinks below. The opposite is true at higher latitudes, with the
westward flow experiencing more cooling and sinking below the eastward
flow. This cooling trend can be seen in the lower panel of Figure
(\ref{fig:3d_tph}). Lastly, very little motion is apparent deeper in
the planet, as the flows are confined to a relatively small region
near the top of the planet, supporting our assumption of a spherically
symmetric inner boundary condition.

\subsubsection{Sub-surface Thermal Stratification}
\begin{figure}
\includegraphics[height=8.0cm,width=9.0cm]{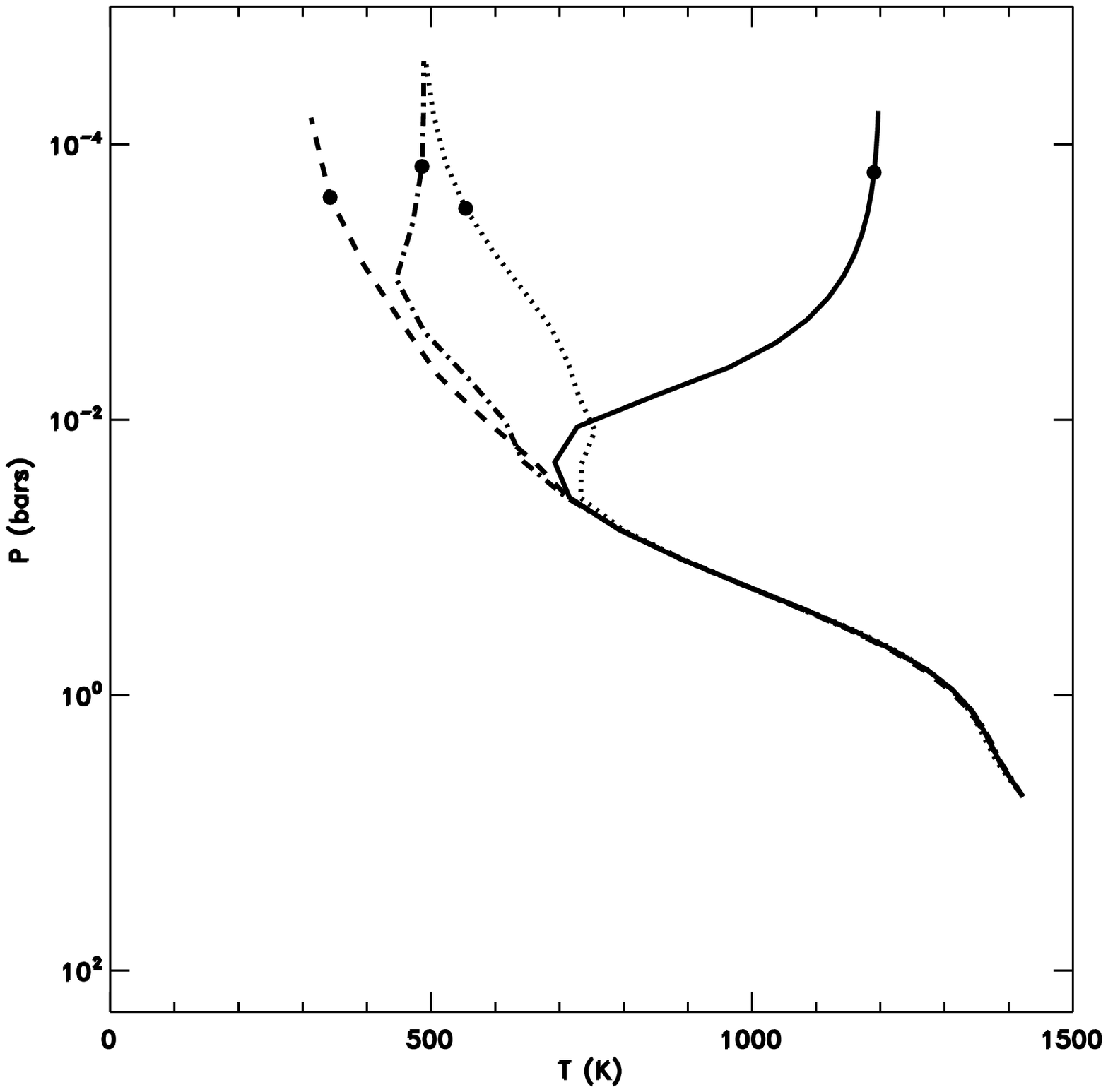}
\includegraphics[height=8.0cm,width=9.0cm]{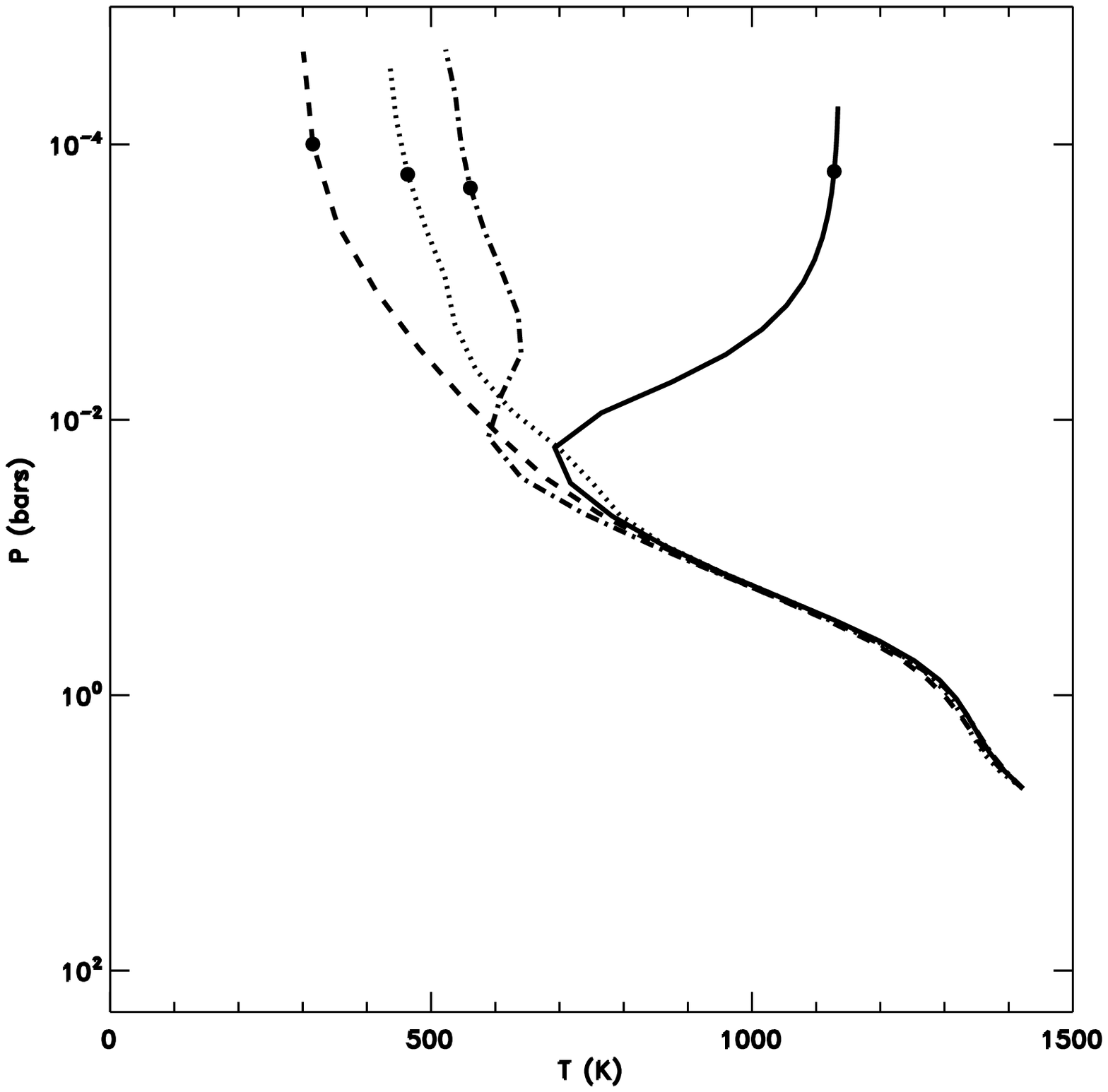}
\caption{The pressure temperature profiles within a planet spinning
with a $3 \mathrm{day}$ period. The left-hand panel shows profiles at
the equator ($\theta=0$), while the right-hand panel shows profiles
from a latitude of $\theta=35^{\circ}$. The individual lines
represent $\phi=0$ (solid line), $\phi=\frac{\pi}{2}$ (dotted),
$\phi=\pi$ (dashed), and $\phi=\frac{3\pi}{2}$ (dash-dot). The dot on
each profile denotes the location of the photosphere.}
\label{fig:3d_PT}
\end{figure}

Another method of visualizing the structure of the planet is through
pressure-temperature profiles. Of crucial importance when calculating
the emergent spectra, the pressure-temperatures at four different
longitudes are shown in Figure (\ref{fig:3d_PT}) for both the equator
and $\theta=35^{\circ}$. All profiles agree well at pressures above
$0.1 \mathrm{bar}$, again supporting the assumption of spherical
symmetry at depth.  However, below this pressure (or above this
height), their behaviors are quite different.

The day-side profile undergoes a significant temperature inversion
({\it i.e.} temperature begins to increase with radius and decrease
with pressure). This temperature gradient allows radiative diffusion
from the photosphere to the planetary interior. This excess flux is
advected to the night-side deep down in the planetary envelope.
Comparing to Figure (\ref{fig:3d_teq}), it is evident that the lowest
temperature region, $\sim 650\mathrm{K}$, is associated in part with
the cool, return flow. This transitional region is analogous to the
thermocline in the terrestrial ocean which separates the surface and
deep water layers.

Near the upper atmosphere (where the pressure is low), the temperature
distribution contains a large, approximately isothermal region on the
day-side, clearly dominated by the stellar irradiation. In contrast,
the temperature of the night-side ($\phi=\pi$) monotonically decreases
throughout the entire atmosphere, reaching a temperature of $\sim 300
\mathrm{K}$ at the photosphere. While the day-side is fully radiative,
there is a region near the photosphere on the night-side that is
convectivly unstable in the radial direction.  As a consequence of
efficient convective transport, the $P-T$ distribution on the night
side is approximately adiabatic.

The profiles near the terminators (at both $\phi=\frac{\pi}{2}$ and
$\phi=\frac{3\pi}{2}$) exhibit more complex structures due to
considerable differences in the advective transport of heat. They also
show differences between the equatorial values (left-hand panel) and
higher latitudes (right-hand panel). At the equator, the temperature
at $\phi=\frac{\pi}{2}$ exhibits an isothermal region ranging from
$0.1$ to $10^{-3}$ bars. This terminator is associated with the
prograde flow. By the time this flow reaches $\phi=\frac{3\pi}{2}$ it
has cooled substantially. A slightly hotter region is evident from the
westward flow at lower pressures (larger radius) near the
photosphere. For the profiles from $\theta=35^{\circ}$, it is at
$\phi=\frac{3\pi} {2}$ where an approximately isothermal region exists
at depth. The $\phi=\frac{\pi}{2}$ profile at high latitudes decreases
monotonically, as very little heat is advected eastward.

\subsection{Opacity Effects}
\label{sec:opacity}
The effect of opacity can not be understated. Opacities regulate the
efficiency of both the absorption of the incident stellar irradiation
on the day-side and the re-radiation from the night-side. As noted by
previous authors, major uncertainties in composition, metallicity, and
chemistry all cause significant changes to the opacity. Furthermore,
the inclusion of clouds, characterized by models of particle size
distributions and vertical extent, tends to smooth out wavelength
dependent opacities, resulting in spectral energy distributions that
more closely approximate blackbodies. In our current models, opacities
are found using the tables of \citet{pollack1985} for lower
temperatures coupled with \citet{alexander1994} for higher
temperatures. These are Rosseland mean opacities and include the
effects of atomic, molecular, and solid particulate absorbers and
scatters.

\begin{figure}
\begin{center}
\includegraphics[height=8.0cm,width=12.0cm]{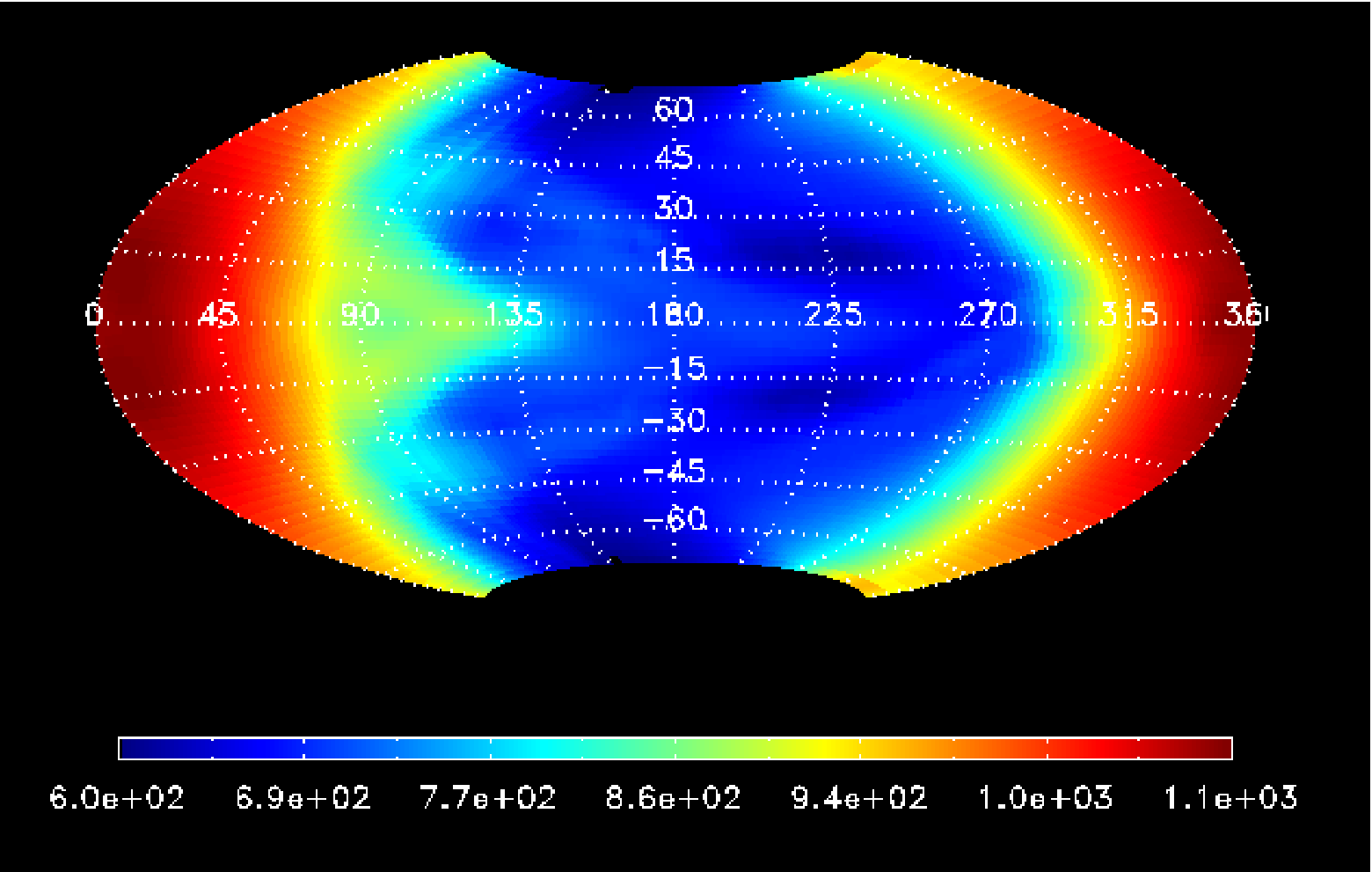}
\includegraphics[height=8.0cm,width=12.0cm]{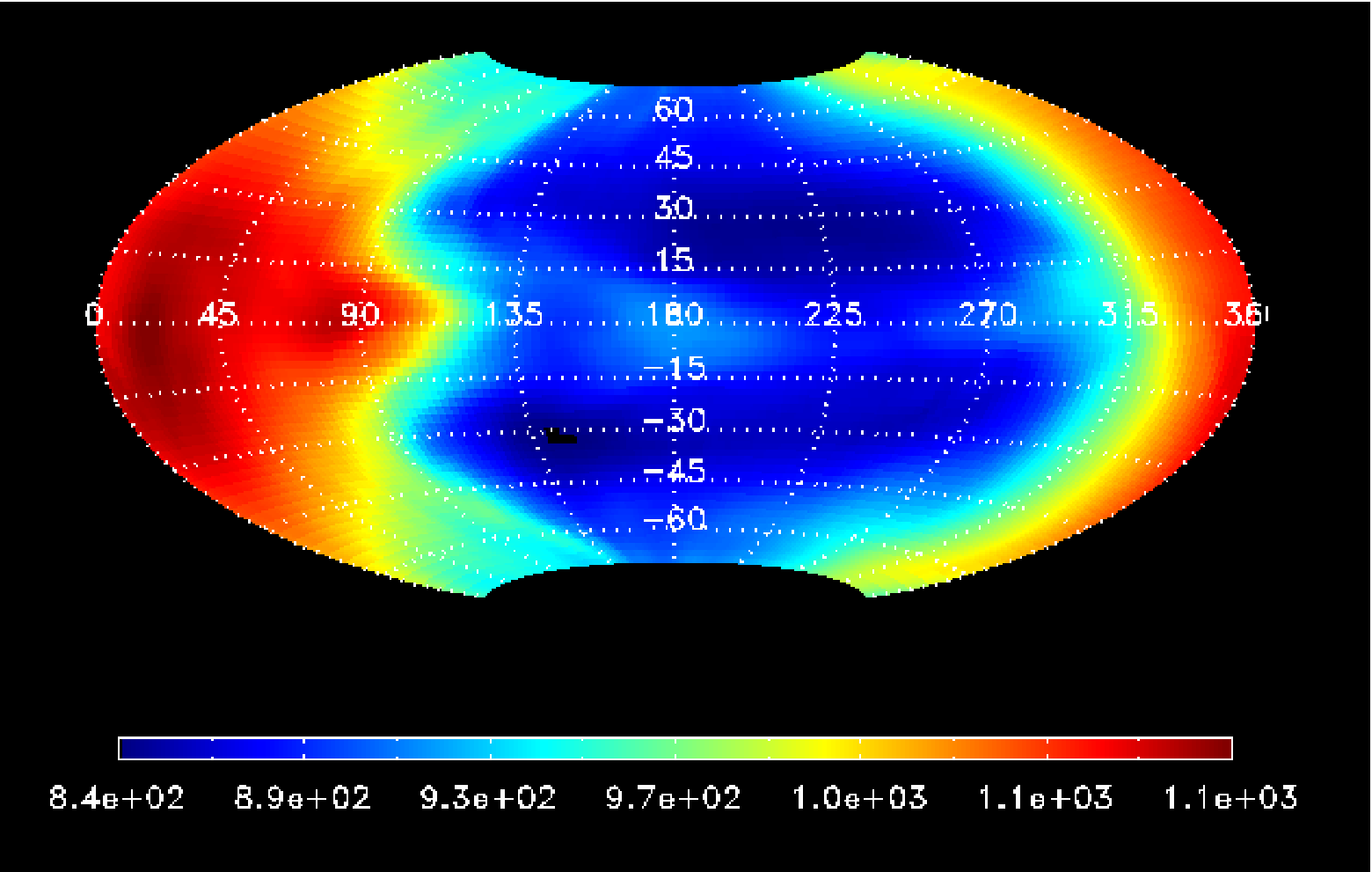}
\end{center}
\caption{The temperature at the photosphere of a model with the
opacities reduced by a factor of $100$ (upper panel) and $1000$ (lower
panel). Lower opacity fluid absorbs the incident stellar irradiation
deeper in the atmospheres. This higher density material is able to
advect the energy to the night-side more efficiently, leading to larger
night-side temperatures.}
\label{fig:T_opc_ph}
\end{figure}

In light of the uncertainties associated with opacity, we have studied
the effect of uniformly changing the opacity by some multiplicative
factor. More detailed studies of specific, temperature and density
dependent augmentations to the opacity will be presented
elsewhere. Figure (\ref{fig:T_opc_ph}) shows the temperature
distribution across the photosphere of a planet with opacities reduced
by both a factor of $100$ (upper panel) and a factor of $1000$ (lower
panel). In comparison to Figure (\ref{fig:3d_tph}), night-side
temperatures are significantly higher for both models with lower
opacities, and distributions are smoother. In addition, due to changes
in the flow detailed below, the hottest spot is displaced slightly
from the sub-solar point. At the equator, the displacement is $\sim
10^\circ$ and $\sim 20^{\circ}$ eastward (in the direction of
rotation) for opacity reductions of $100$ and $1000$ respectivly. This
displacement is largest at the equator, with maximum temperatures at
higher/lower latitudes occurring closer to the sub-solar longitude.

In a previous analysis (\citet{burkert2005}), we derived a formula for
the night-side temperature by equating the radiative timescale with
the crossing timescale.  This two-zone (day-night) model assumes that:
the advective heat flux is larger then the heat flux from the
interior, the heat carried by a day-night thermal current is
determined by the amount of radiative diffusion during the hemispheric
circulation, and the night-side radiates all the heat advected to its
proximity as a black body. With these assumptions, the night-side
temperature can be estimated by
\begin{equation}
T_n = \left(\frac{4vc_d^2}{3\pi\kappa_d\sigma R_p}\right)^{1/4}
\label{eq:Tburkert}
\end{equation}
which is an decreasing function of opacity. In this formula, $v$ is
the average advective speed which is on the order of the day-side
sound-speed $c_d$, and $\kappa_d$ is the day-side opacity. The
increase of night-side temperature with decreasing opacity reflects
the depth that incident stellar irradiation is deposited on the
day-side. If the atmosphere contains grains with an abundance and size
distribution comparable to that of the interstellar medium, only
shallow heating occurs on the day-side, and the circulation does not
effectively transmit heat to the night-side, which then cools well
below the day-side. However, as the abundance of grains in the
atmosphere is reduced, the stellar radiative flux penetrates more
deeply into the atmosphere on the day-side, and the higher density
atmospheric circulation carries a larger flux of heat over to the
night-side.

For the parameters used here, equation (\ref{eq:Tburkert}) predicts
$T_n \sim 250\mathrm{K}$ for the interstellar opacity simulation,
while for the lower opacity simulations, the formula predicts $\sim
750\mathrm{K}$ and $\sim 950\mathrm{K}$ for reductions of $100$ and
$1000$ respectively. Inspection of Figures (\ref{fig:3d_tph}) and
Figure (\ref{fig:T_opc_ph}) show average night-side temperatures of
$\sim 380\mathrm{K}$, $\sim 700\mathrm{K}$, and $\sim 890\mathrm{K}$
for the same three cases. These results clearly indicate that the
night-side temperature decreases with the magnitude of the opacity.
The predicted values from equation (\ref{eq:Tburkert}) generally agree
with those from the simulation and it provides a framework for
understanding the global heat flow. The differences in these results
can be attributed to several factors; the surface heat flux carried by
the thermal current from the day to night-side is not entirely
radiated on the night-side, but rather cools as it travels and
advection at depth plays an important role in transporting heat. These
effects can be incorporated into a more comprehensive four-zone
(day-night and interior-photosphere) model where we examine the energy
transfer within the optically thick regions below the planetary
photosphere.

\begin{figure}
\begin{center}
\includegraphics[height=8.0cm,width=12.0cm]{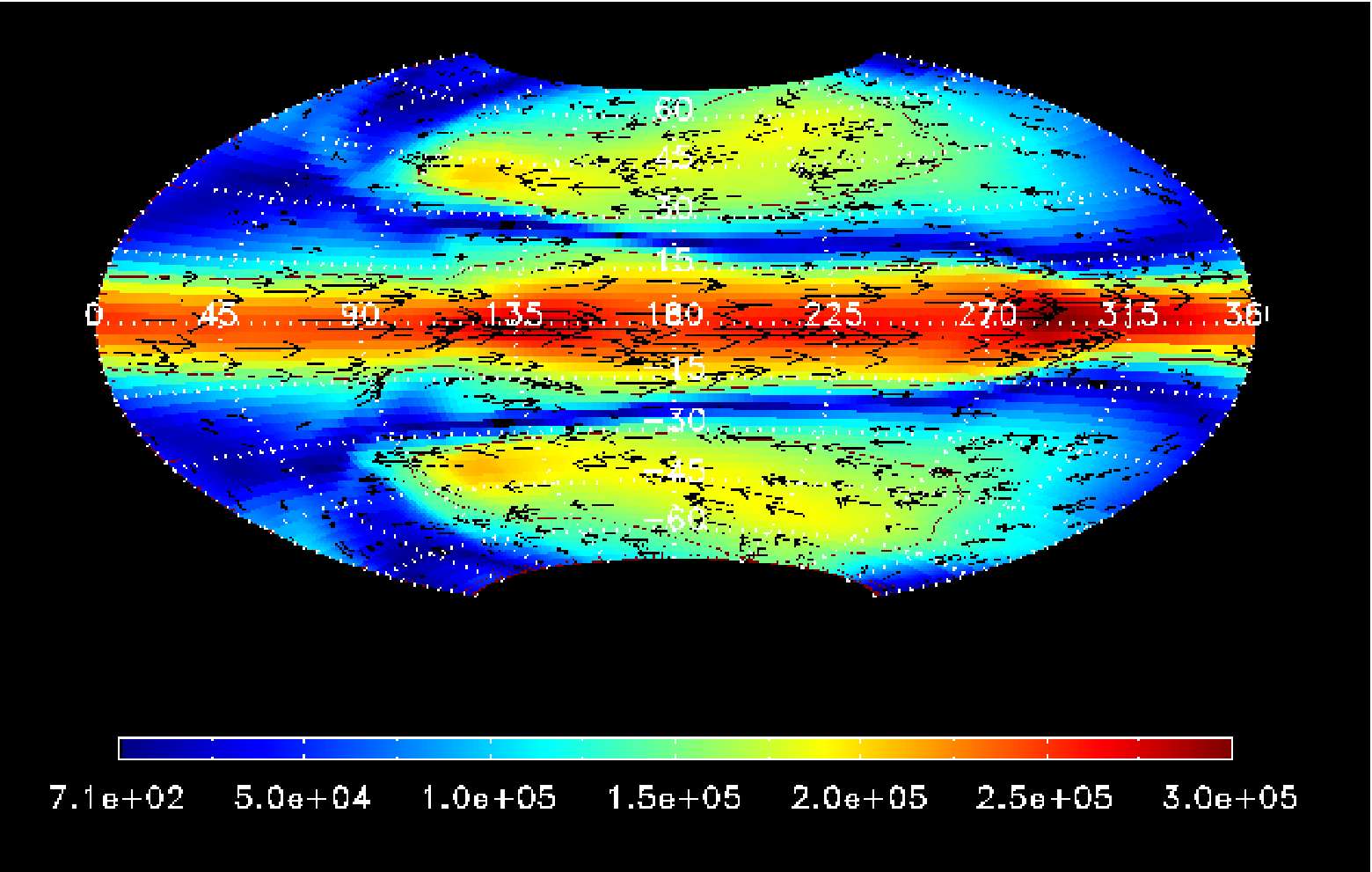}
\includegraphics[height=8.0cm,width=12.0cm]{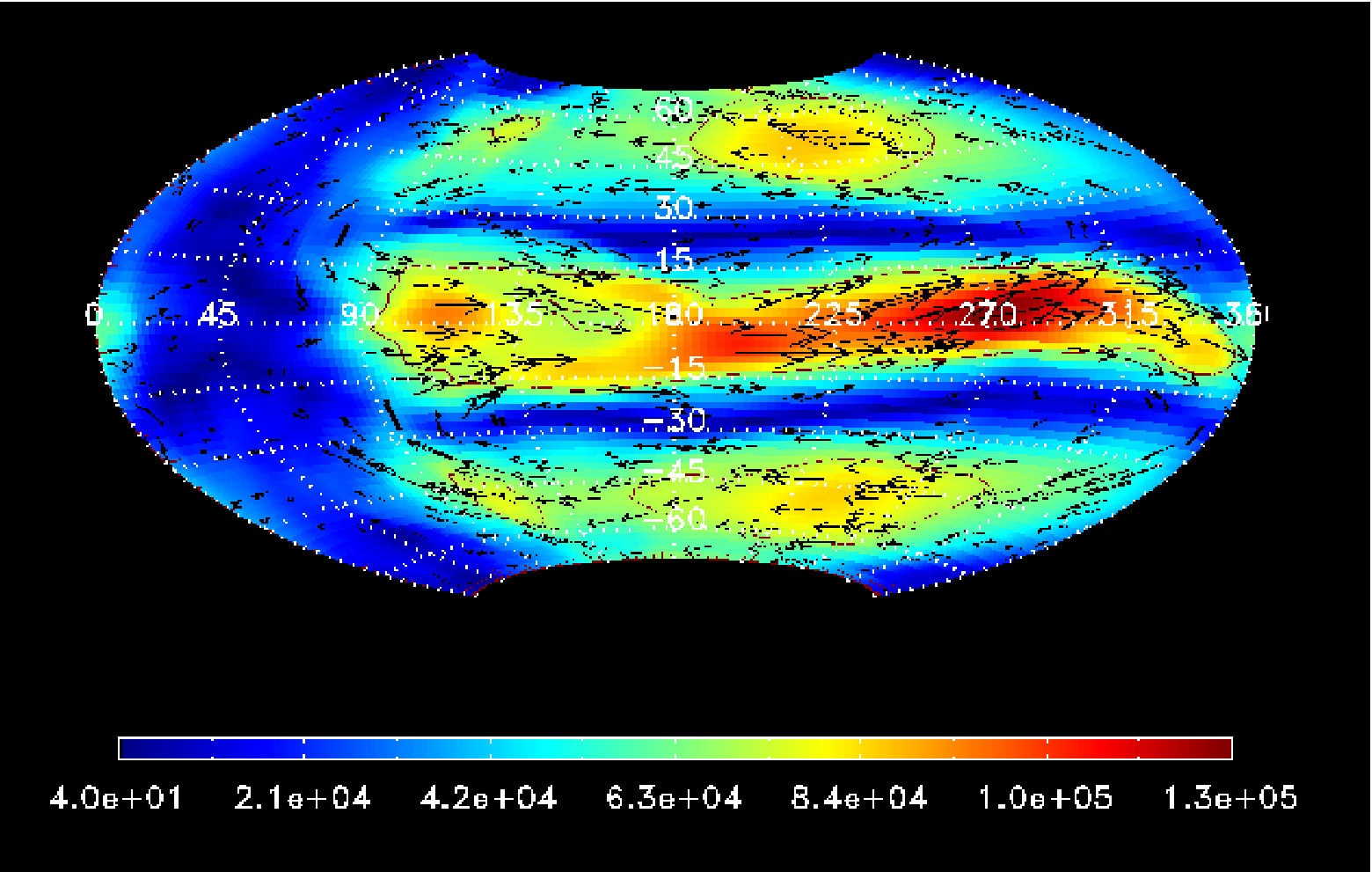}
\end{center}
\caption{Velocity at the photosphere of models with opacities reduced
by a factor of $100$ (upper panel) and $1000$ (lower
panel). Velocities, determined by both the day-night temperature
differential and the cooling efficiency, decrease with deceasing
opacity.}
\label{fig:vph_opc_ph}
\end{figure}

Before the presentation of the four-zone model, it is useful to
analyze the results of the numerical simulation.  These calculations
indicate that changing the opacity of the atmosphere not only alters
the night-side temperature, but it also modifies both the flow
dynamics and interior structure. For large opacity (our standard
case), the isothermal region on the day-side is relatively shallow,
and thus the increased cooling leads to lower night-side
temperatures. The large temperature differential promotes a fast flow
velocities around the planet. As the opacities are decreased, and the
night-side temperature increases and the velocities decrease.  In the
lowest opacity simulation, the velocity remains subsonic throughout
the entire simulation. A more uniform temperature across the planet
surface also allows for circumplanetary flow near the equator by
reducing the pressure gradient. Figure (\ref{fig:vph_opc_ph}) shows
the velocity at the photosphere of the two low-opacity
simulations. These results should be compared to the standard-value
results shown in Figure (\ref{fig:v_coriolis_ph}). As in the standard
case, flow at higher and lower latitudes still travels westward at the
$\frac{3\pi}{2}$ terminator and circumplanetary flow at the surface is
suppressed due to increased cooling times.

\begin{figure}
\includegraphics[height=8.0cm,width=9.0cm]{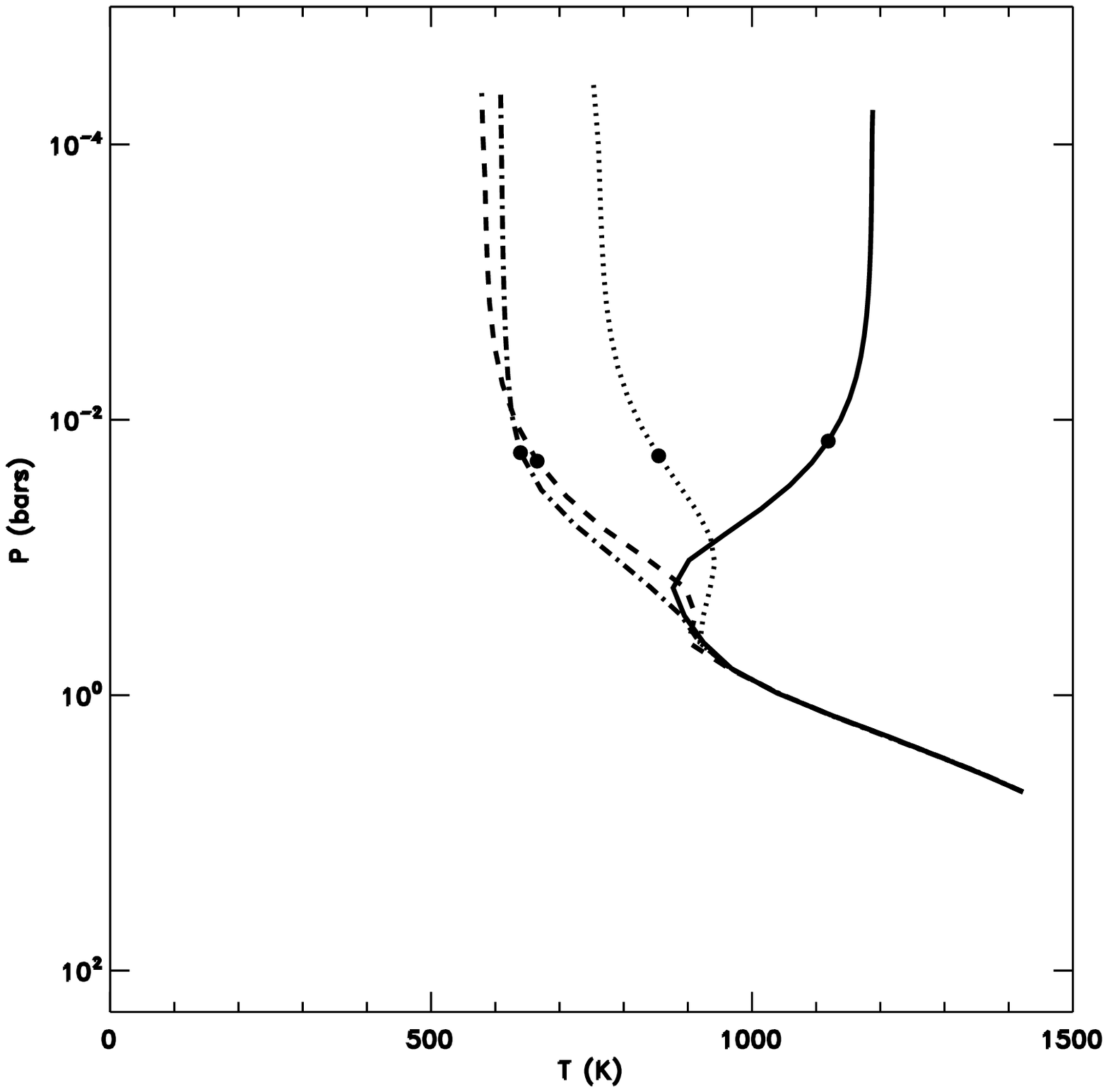}
\includegraphics[height=8.0cm,width=9.0cm]{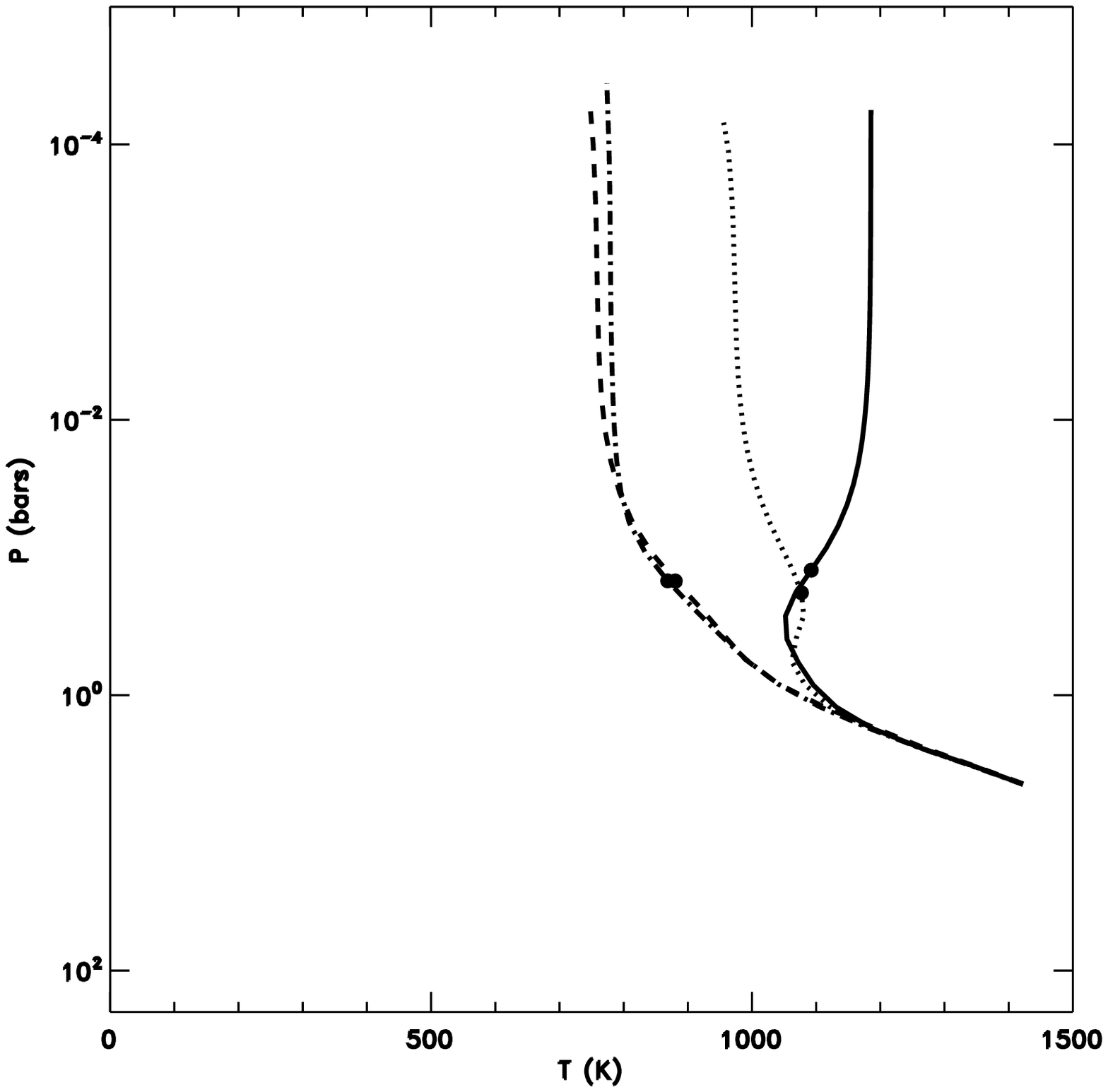}
\caption{The pressure temperature profiles at the equator of a planet
spinning with a $3 \mathrm{day}$ period. The left-hand panel shows
profiles for the run with opacities reduced by a factor of $100$,
while the right-hand panel shows profiles for a simulation with
opacities reduced by a factor of $1000$. The individual lines
represent $\phi=0$ (solid line), $\phi=\frac{\pi}{2}$ (dotted),
$\phi=\pi$ (dashed), and $\phi=\frac{3\pi}{2}$ (dash-dot). The dot on
each profile denotes the location of the photosphere, considerably
deeper then in the standard opacity simulation.}
\label{fig:lowopc_PT}
\end{figure}

In Figure (\ref{fig:lowopc_PT}) we show the pressure-temperature
profiles at the equator for the two reduced opacity simulations. In
contrast to Figure (\ref{fig:3d_PT}) where the night-side was fully
convective, both lower opacity simulations exhibit isothermal regions
in the upper atmospheres around the entire planet. Below the
photosphere, there is a slightly negative temperature gradient so that
the day-night advective heat flux deep beneath the photosphere can
radiatively diffuse to the planet's photosphere. As it is to be
expected, the radial extent of the nearly isothermal atmosphere is
largest with the lowest opacity. Both simulations retain a convective
regions below the isothermal regions on the night-side.

\subsubsection{A Four-Zone Model for the Sub-Surface Thermal Structure of Planetary Atmospheres}
Discussions in the previous section clearly indicate that thermal
currents not only transport heat from the day to night, but also from
the night to day-side at other radii. In addition, the day-side
atmosphere is radiative and isothermal, while the night-side is
convective and adiabatic. Finally, quasi hydrostatic equilibria is
maintained in most regions except the subduction zones which separate
the thermal currents.  We now take these effects into consideration in
the determination of the night-side temperature and the sub-surface
thermal structure.

The two-zone approximation in equation (\ref{eq:Tburkert}) is based on
the assumption that the radiative and crossing time-scales in the
planetary atmosphere are nearly equal. We show in the next section
(Figure (\ref{fig:3d_rad_cross})) that this assumption is only
marginally satisfied on the night-side near the photosphere.  In the
standard opacity model, the radiative timescale around the terminators
is several orders of magnitude shorter than the crossing
timescale. This allows for significant cooling, yielding much lower
night-side temperatures then would be expected from equation
(\ref{eq:Tburkert}). Similar behavior is seen in the lower opacity
models.

In order to analytically account for the transport of heat well below
the photosphere on either side of the planet, we divide it into four
zones, representing the day-night and interior-photosphere regions.
There are then 4 sets of thermodynamic variables including: $P_{DS}$,
$P_{DI}$, $P_{NS}$, $P_{NI}$, $T_{DS}$, $T_{DI}$, $T_{NS}$, and
$T_{NI}$ where $P$ and $T$ are pressure and temperature, and the
subscripts $D$, $N$, $S$, and $I$ represent day, night, photospheric
surface and interior respectively. There are three additional
quantities which connect these regions: the velocity between the day
and night interiors, $v_{adv}$, the radial distance between the
photospheric surface and the planetary interior, $l_D$, and the
density of the gas at the planetary interior on the day-side
$\rho_{DI}$. These quantities can be solved simultaneously with the
following 11 equations.

Although the total stellar incident flux $F=L_\ast/4 \pi D^2$ (where
$L_\ast$ is the stellar luminosity and $D$ is the distance between the
star and the planet), irradiates only on the day-side, the condition
for thermal equilibriums implies that
\begin{equation}
2 \sigma (T_{DS}^4 + T_{NS}^4)  = F.
\label{eq:4zone1}
\end{equation}
Hydrostatic equilibrium at the photosphere of both the day and night
sides gives:
\begin{equation}
P_{DS} = \frac{2}{3}\frac{g}{\kappa_{DS}}, \ \ \ \
P_{NS} = \frac{2}{3}\frac{g}{\kappa_{NS}}.
\label{eq:4zone2}
\end{equation}
For the simplicity of analytic approximation, we represent opacity as
\begin{equation}
\kappa = \kappa_0 T^\beta.
\label{eq:4zone3}
\end{equation}
In the temperature and density ranges which are relevant to the
present investigation, the opacity is roughly independent of density,
thus we adopt the approximation $\kappa_0 = 0.0391$ and $\beta =
0.641$.

Our numerical results indicate that the pressure may change by more
then two orders of magnitude between the planetary surface and the
temperature inversion layer. In contrast, the temperature changes by
less than a factor of two. In the spirit of analytic simplicity, we
adopt an isothermal approximation for the hydrostatic envelope when we
determine the pressure at the planetary interior on the day-side, such
that
\begin{equation}
P_{DI} = P_{DS}exp\left[\frac{l_d g}{c_{s,d}^2}\right],
\label{eq:4zone4}
\end{equation}
where $c_{s,d} = (\gamma{\sf R}_{G} T_{DS}/\mu)^{1/2}$ is the sound
speed.  However, to calculate the temperature variation, we adopt a
radiative diffusion approximation in the radial direction. Including
the azimuthal advective transport, we have
\begin{equation}
1 - \left( \frac{T_{DI}}{T_{DS}} \right)^{4-\beta} \simeq \left(4 -
\beta\right) \left[ \left(1 - {\rho_{DI} v_{adv} c_{s, d}^2 \over
F}\right) exp\left[\frac{g l_d}{c_{s,d}^2}\right] - 1 \right] \left( 1
+ {T_{NS}^4 \over T_{DS}^4} \right),
\label{eq:4zone5}
\end{equation}
where 
\begin{equation}
\rho_{DI} = \mu P_{DI} / {\sf R}_{G} T_{DI}.
\label{eq:4zone6}
\end{equation}
We choose $l_D$ to be the depth where advection carries half of the
incident flux, {\it i.e.} where
\begin{equation}
\rho_{DI} v_{adv} c_{s, d}^2 \simeq F/2.
\label{eq:4zone7}
\end{equation}
The advective velocity at depth is driven by the pressure
differential, such that
\begin{equation}
v_{adv}^2 \simeq \frac{2 {\sf R}_{G} T_{DI}}
{\mu}ln\left(\frac{P_{DI}}{P_{NI}}\right).
\label{eq:4zone8}
\end{equation}
We assume the consequence of the advective transport is to thoroughly
mix the gas so that planetary interior becomes isothermal with
\begin{equation}
T_{NI} = T_{DI}
\label{eq:4zone9}
\end{equation}

For the standard opacity model, the night-side remains fully
convective. Assuming convection is efficient, the envelope of the
night-side is expected to be adiabatic, such that
\begin{equation}
\frac{T_{NI}}{T_{NS}} = \left(\frac{P_{NI}}{P_{NS}}\right)^{\frac{\gamma-1}
{\gamma}}.
\label{eq:4zone10}
\end{equation}
The energy equation for the night-side can be written as
\begin{equation}
\left(1-\frac{P_{DI}}{P_{NI}}\right) = -\frac{v_{conv}}
{v_{adv}}\frac{g\mu\pi R_p}{{\sf R}_{G} T_{NI}}
\label{eq:4zone11}
\end{equation}
where the convective velocity $v_{conv}$ can be obtained from the
mixing length theory. For the low opacity models, the upper regions of
the night-side become stabilized against convection. We can then
replace the energy equation with the diffusion approximation and
recalculate the thermal structure accordingly.

With these equations, we can provide a set of algebraic equations
which essentially reproduce the behavior of our numerical
simulation. These equations are also more comprehensive than that in
equation (\ref{eq:Tburkert}). For opacity similar to that of the
interstellar medium ({\it i.e.} no reduction in $\kappa_0$), the
night-side remains convective so that
equations(\ref{eq:4zone1})-(\ref{eq:4zone11}) are valid. In this
limit, $T_{DS}$ is sufficiently larger than $T_{NS}$ such that
equations (\ref{eq:4zone1}) and (\ref{eq:4zone2}) lead to complete
information on the surface layer of the day-side,
\begin{equation}
T_{DS}^4 = F/2 \sigma, \ \ \ \ \ \ P_{DS} = 2 g / 3 \kappa_0 T_{DS}^\beta.
\end{equation}
In comparison with the large variations in equation (\ref{eq:4zone4}),
the day-side is nearly isothermal with $T_{DI} \sim T_{DS}$. At the
depth where half of the incident stellar flux is advected from the day
to night-side, equation (\ref{eq:4zone7}) yields
\begin{equation}
P_{DI} v_{adv} \simeq {F \over 2}, \ \ \ \ \ l_d \simeq {c_{s, d} ^2 \over g}
{\rm ln} \left( {3 F \kappa_0 T_{DS}^\beta \over 4 g v_{adv} }\right).
\end{equation}
In the fully adiabatic night-side, we find from equations
(\ref{eq:4zone2}), (\ref{eq:4zone4}), (\ref{eq:4zone8}), and
(\ref{eq:4zone10}) that
\begin{equation}
P_{NS} = P_{DS} (T_{NI}/T_{NS})^\beta,
\end{equation}
and
\begin{equation}
{T_{NS} \over T_{DS}} = \left[ {\rm exp} \left(
{ v_{adv}^2 - 2 g l \over 2 c_{s, d}^2 } \right) \right]^{(\gamma-1)/
(\gamma+ \beta (\gamma-1))}.
\label{eq:4zone12}
\end{equation}
From equations (\ref{eq:4zone8}), (\ref{eq:4zone9}), and
(\ref{eq:4zone11}), we find
\begin{equation} 
{\rm exp}\left[{v_{adv}^2 \over 2 c_{s, d} ^2}\right] - {v_{conv} \over v_{adv} }
{g \pi R_p \over c_{s, d}^2 } = 1. 
\label{eq:4zone13}
\end{equation}
In the limit that $v_{adv} < c_{s, d}$, equation (\ref{eq:4zone13}) reduces to
\begin{equation}
v_{adv} \sim ( 2 v_{conv} g \pi R_p) ^{1/3}.
\end{equation}
The convective speed can be estimated from the mixing length theory
which is mostly determined by $T_{NI} \sim T_{DS}$ and $\rho_{NI}$
These quantities have very little or no dependence on $\kappa_0$ so
that $v_{adv}$ does not change significantly as a function of
$\kappa_0$. In this limit, equation (\ref{eq:4zone12}) becomes
\begin{equation}
{T_{NS} \over T_{DS}} \propto \left[ {4 g v_{adv} \over 3 F \kappa_0
T_{DS}^\beta} \right]^{((\gamma-1) -1)/ (\gamma+ \beta (\gamma-1))}
\propto \kappa_0 ^{-1/4}.
\label{eq:4zone14}
\end{equation}
Thus, the four-zone model generates a similar $T_{NS}$ dependence on
$\kappa_0$ as equation (\ref{eq:Tburkert}). In comparison with
numerical results, equation (\ref{eq:4zone14}) does reasonably well
for high opacity convective simulations, but expression would be
improved by considering the possibility that the night-side may also
become radiative in the limit of very low $\kappa_0$.

\section{Analysis and Model Comparisons}
\label{sec:comparison}
A number of groups, including \citet{showman2002}, \citet{cho2003},
\citet{burkert2005}, \citet{cooper2005}, \citet{cho2006}, and
\citet{langton2007} have carried out non-linear numerical simulations
studying the dynamics of hot-Jupiter atmospheres. Both the methods and
results vary considerably. In this section, we attempt to compare
their methodology to that presented here, concentrating on the
assumptions that lead to differing results.

Both \citet{showman2002} and \citet{cooper2005} solve the primitive
equations, the former using the EPIC code developed by
\citet{dowling1998} and the later using the ARIES/GEOS Dynamical Core
model initially developed by \citet{suarez1995}. The primitive
equations, cast in isobaric coordinates, are widely used in
meteorology. When deriving them from the full Navier-Stokes equation,
a key assumption of hydrostatic equilibrium is invoked. Assumption of
hydrostatic equilibrium explicitly links the thickness of a layer to
its local temperature. Radial motion can only occur in conjunction
with divergence along isobars, the magnitude determined by that which
maintains the condition of hydrostatic equilibrium. With this
assumption, the continuity equation and hydrostatic condition take the
place of the radial momentum equation, and can be written as
\begin{equation}
\nabla_p {\bf \cdot v} + \frac{\partial \omega}{\partial p} = 0,
\end{equation}
where ${\bf v}$ is the horizontal velocity and $\omega$ is the
vertical velocity with respect to pressure coordinates. The resulting
radial velocity is slow in comparison to the horizontal (or isobaric)
motion. Although an excellent assumption for thin terrestrial
atmospheres, it neglects critical flows present in thicker atmospheres
such as those of hot-Jupiter's. Although deviation from hydrostatic
equilibrium provides the dominate radial acceleration, high velocity
azimuthal flows on the night-side give rise to non-negligible radial
forces neglected in the primitive equations that contribute to the
ability of the converging flows to pass under one another as seen in
Figure(\ref{fig:3d_veq_high}).

\begin{figure}
\begin{center}
\includegraphics[height=8.0cm,width=12.0cm]{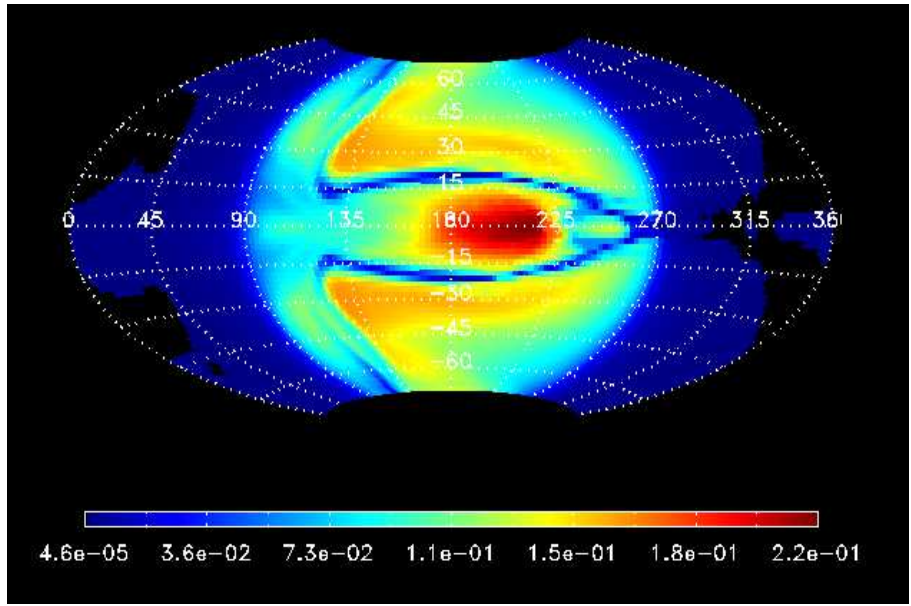}
\end{center}
\caption{The ratio of radiative timescale ($\tau_{rad}$) to the
crossing timescale ($\tau_{x}$) at the photosphere of the simulation
with interstellar opacities. Near the terminator, the primary region
of interest, the radiative timescale is $\sim 5$ orders of magnitude
shorter then the crossing timescale. This rapid cooling accounts for
the sharp temperature gradient near the terminator seen in Figure
(\ref{fig:3d_tph}).}
\label{fig:3d_rad_cross}
\end{figure}

Our numerical results and analytic approximation clearly indicate
the dependence of the night-side temperature on the efficiency
of radiation transfer.  The heat flux carried by the thermal current
is determined by the penetration depth of the incident stellar radiation.
Radiative diffusion and convection are essential processes which 
regulate the heat diffusion well below the photosphere of the planet.
In conjunction with the primitive equations, both \citet{showman2002}
and \citet{cooper2005} utilize a Newtonian radiative scheme to
approximate stellar irradiation. In principle, this scheme is only
appropriate for energy deposition and emission in optically thick 
regions. The Newtonian radiative scheme
provides a heating/cooling term to the energy equation that relaxes
the temperature toward some pre-defined equilibrium distribution on
some radiative timescale. The forcing term can be expressed
\citep{cooper2005} as,
\begin{equation}
\frac{q}{c_v} =
-\frac{T\left(\theta,\phi,p,t\right)-T_{eq}\left(\theta,\phi,p\right)}{\tau_{eq}\left(p\right)}.
\end{equation}
For the radiative timescale, $\tau_{eq}$, \citet{showman2002} assume a
constant value given by that at $5 \mathrm{ bars}$, while
\citet{cooper2005} use the calculations of \citet{iro2005} to set a
radiative relaxation timescale that is dependent on the local
pressure. As noted by the authors, this approximation is crude, but
allows rapid computation of a large number of models. In order for the
Newtonian approximation to be viable, the radiative timescale
($\tau_{rad}\approx\frac{E_T}{F}$) must be much longer than the
crossing timescale ($\tau_{x}\approx\frac{\pi R_p}{2
\left|v\right|}$). In this limit, the temperature distribution will be
determined by the dynamics, rather than the assumed equilibrium
distribution. This assumption becomes problematic in the upper
atmosphere. In Figure (\ref{fig:3d_rad_cross}) we plot the ratio of
timescales $\frac{\tau_{rad}}{\tau_{x}}$ at the photosphere of the
planet.

It is clear from Figure (\ref{fig:3d_rad_cross}) that $\tau_{rad} <
\tau_{x}$ throughout the photosphere. The most significant deviation
occurs on the day-side, where there is very little motion, and
$\tau_{rad}$ exceeds $\tau_{x}$ at the planetary photosphere 
by many orders of magnitude. Near the terminator, where the stellar 
irradiation falls drastically, the surface 
radiative timescale is also much shorter then the dynamical timescale;
it is here that the deviations from a Newtonian radiative scheme are
most important. The flow radiates a significant portion of its energy
before it flows to the night-side, allowing for the sharp edges in the
temperature distribution seen in Figure (\ref{fig:3d_tph}), despite
high velocity flows. In contrast, the Newtonian assumption that
$\tau_{rad} >> \tau_{x}$, implies that the fluid carries a significant
quantity of heat, leading to an overall distribution that will be more
uniform with longitude. Finally, despite substantial radiation near
the terminators, the near equality of $\tau_{rad}$ and $\tau_{x}$ at
$\phi=\pi$ demonstrates that the temperature at the back-side is
determined purely by the advection of energy by the flow.

\begin{figure}
\begin{center}
\includegraphics[height=8.0cm,width=12.0cm]{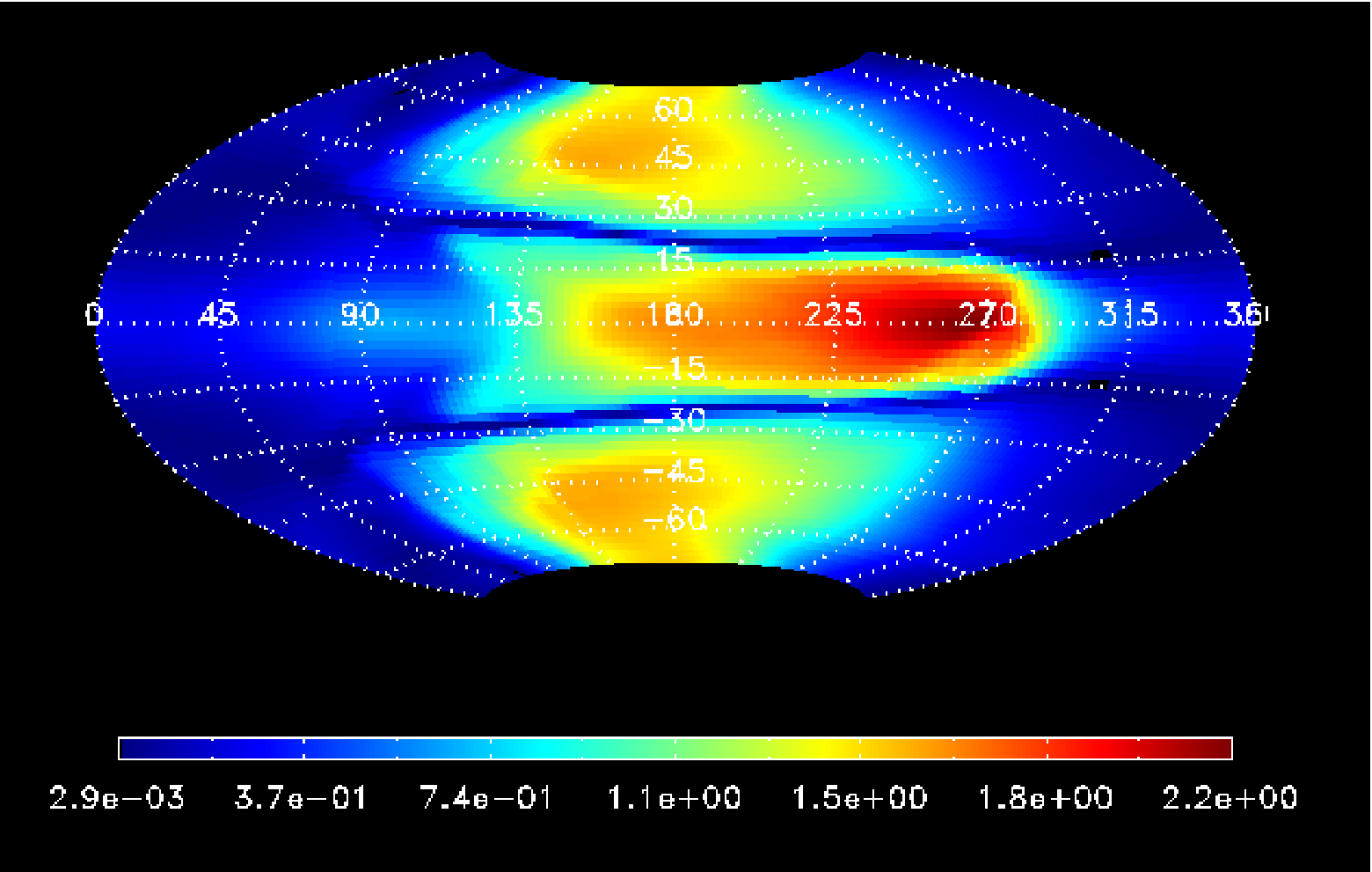}
\includegraphics[height=8.0cm,width=12.0cm]{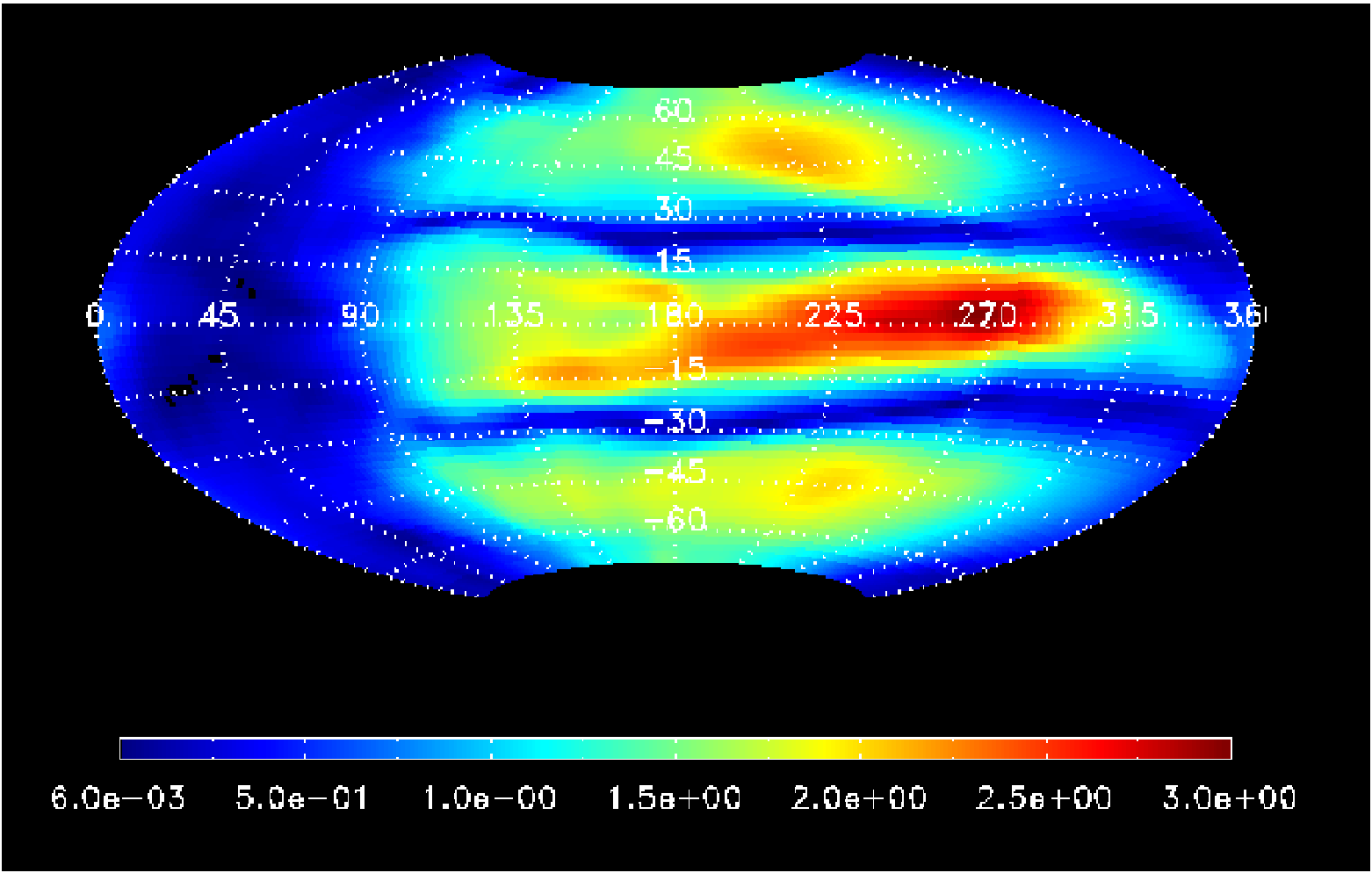}
\end{center}
\caption{The ratio $\frac{\tau_{rad}}{\tau_{x}}$ at the photosphere of
models with opacities reduced by a factor of $100$ (upper panel) and
$1000$ (lower panel). Although lowering the opacity decreases the
ability of the fluid to radiate its thermal energy, the radiative
timescale remains several orders of magnitude shorter then the
crossing timescale for most of the photosphere.}
\label{fig:radtime_opc_ph}
\end{figure}

To again highlight the effect of opacity, Figure
(\ref{fig:radtime_opc_ph}) shows the ratio of radiative and crossing
timescales, $\frac{\tau_{rad}}{\tau_{x}}$, for the runs with varying
opacity. The upper and lower panels show the simulations with
opacities reduced by a factors of $100$ and $1000$
respectively. Again, the most important area in determining the
behavior is near the terminators. In comparison to Figure
(\ref{fig:3d_rad_cross}), it is obvious that lower opacity fluid is
able to advect energy to the night-side more efficiently due to
increased cooling times. As mentioned above, lower opacity on the
day-side, allows heat to penetrate further into the planet, and
re-radiation near the terminator is somewhat suppressed. Although
Figure (\ref{fig:radtime_opc_ph}) shows a marked increase in
$\frac{\tau_{rad}}{\tau_{x}}$ from the interstellar opacity
simulation, the Newtonian approximation ($\tau_{rad} >> \tau_{x}$) is
never satisfied. The largest values, $\frac{\tau_{rad}}{\tau_{x}}\sim
3$, only occur on the night-side in the high velocity circumplanetry
jet.

The studies of \citet{cho2003}, \citet{menou2003}, \cite{cho2006}, and
\cite{langton2007} take a different approach, concentrating on the
effect of eddies and waves on the overall dynamics. \citet{cho2003}
\citet{menou2003} and \cite{langton2007} use the shallow-water
equations, while \cite{cho2006} solves the more generalized
equivalent-barotropic equations. The equivalent-barotropic equations
can be derived by vertically integrating the primitive equations
described above. The resulting equations describe fluid flow in a
single, isentropic layer whose scale height can vary. Stellar heating
is prescribed by a deflection of the scale-height as a function of
position. Concentrating on a single layer allows for the high
resolution necessary to study the effect of turbulent eddies and
waves. However, allowing for only a single layer implies that the
atmosphere is radial isothermal. Although not a bad assumption in the
upper regions on the day-side, the rest of the atmosphere exhibits
significant radial temperature structure. As noted above, cooling in
the upper regions of the night-side allows material to sink radial and
initiate a return flow. Disallowing this flow would significantly
alter the resulting dynamics.

The final dynamical study of hot-Jupiter's was done by
\citet{burkert2005}. The numerical model they used is quite similar to
the model presented in this paper. They solve the full hydrodynamical
equations, given by equations (\ref{eq:continuity}),
(\ref{eq:momentum}) and (\ref{eq:energy}) together with flux-limited
radiation diffusion. However, they restrict their attention to the
$r-\phi$ plane from $\phi=0\rightarrow\pi$, neglect the curvature
terms in equation (\ref{eq:momentum}), and ignore the effects of
rotation. Comparing our non-rotating results to \citet{burkert2005}
'Case 1' with our 'standard' opacity, we find that the two models
agree quite well. The only substantial differences are that our
backside temperature is slightly lower ($\sim 415 \mathrm{K}$ compared
to their $480 \mathrm{K}$), and the convergence point is seen to
oscillate in the simulations presented here. Our cooler temperatures
can easily be explained by considering the increased compressional
heating that the fluid in \citet{burkert2005} simulations experiences
as it hits the boundary at $\phi=\pi$, and our extension into
three-dimensions. Flows that are able to spread in latitude will cool
more then those confined to the equator. Given that our non-rotating
simulation results in two symmetric convective cells suggests that by
restricting attention to $\phi=0\rightarrow\pi$, they did not miss any
fundamental physics for that scenario. However, the addition of
rotation significantly changes flow patterns and allows for increased
cooling.

In summary, four fundamental differences are included in the models
presented here, that are not included in some way in the previous
methods: self-consistent radiative transfer, solution of the full
radial momentum equation, $3$-dimensions, and rotation. Noticeable
changes occur due to the inclusion of each. Self-consistent radiative
transfer allows significant cooling as the fluid travels around to the
terminators and night-side of the planet. Coriolis forces alter the
structure of the flows, alternately compressing and diverging
streamlines moving around the planet in different directions, and the
multi-dimensional aspect changes the resulting temperature
distribution. Finally, the full treatment of the radial momentum
equation yields significant, non-hydrostatic radial structure and
motion. Disallowing efficient cooling, and limiting radial motion,
while subjecting the planet to the continual stellar energy input,
will lead to much more uniform temperature distributions across the
entire planet then those that we observe here.

\section{Discussion}
\label{sec:discussion}
In this paper we have considered the flow dynamics and heat
redistribution in a tidally locked hot-Jupiter orbiting a solar-type
star with an orbital period of $3 days$. We utilized a
three-dimensional model that solves the full hydrodynamical equations
and models radiative transfer the flux-limited radiation diffusion. We
show that the temperature distribution across the planetary atmosphere
is a sensitive function of its opacity. Our models exhibit strong
day-night temperature contrasts, despite strong winds, the size of
which increases with increasing atmospheric opacity. Large temperature
differences are due, in a large part, to significant cooling of the
flow near the terminators of the planet. Rotational effect
significantly alter the flow pattern, allowing for increased cooling
and suppressing surface circumplanetary flow in the higher opacity
atmospheres.

In our standard model, opacities are calculated using the tables of
\citet{pollack1985} for lower temperatures coupled with
\citet{alexander1994} for higher temperatures. These are interstellar
Rosseland mean opacities and include the effects of atomic, molecular,
and solid particulate absorbers and scatters. Although there is
significant room for improvement in our treatment of opacity, we
explore the effect of uniformly reducing them by factors of $100$ and
$1000$. In agreement with the two-dimensional simulations of
\citet{burkert2005}, we find that lower opacities yield higher
night-side temperatures. Lower opacities allow incident stellar
irradiation to be deposited at larger depths on the day-side,
decreasing the cooling rate as this energy is advected to the
night-side, and increasing night-side temperatures. The relation
between the night-side effective temperature and the opacity in the
planetary atmosphere is verified by a more comprehensive four-zone
analytic model which takes detailed account of the effect of radiation
transfer on both the day and night-side of the planet. This model
highlights the importance of appropriately treating radiation transfer
in the simulation of atmospheric dynamics on close-in gas giant
planets.

The actual value of opacity in the atmospheres of these planets is
highly uncertain. However, the day-side temperature of short period
gas giant planets around solar type stars is $T_{DS} \sim 1200$K,
which is below the grain destruction temperature. We note that the
flow returning to the day-side is dredged up from a cooler interior
($T_{DI} < T_{DS}$) at the sub-solar point. Thus, throughout the
thermal circulation, refractory magnesium-rich grains are
preserved. Other species of less refractory silicate grains will
sublimate near the planetary photosphere on the day-side, condense on
night-side surface, and be carried along with the returning current
beneath the surface. These thermal currents become turbulent as they
generate specific vorticity and excite instabilities. Small grains are
well coupled to the gas and turbulence induces them to collide and
coagulate. In a static atmosphere, particles with sizes $s_p$ and
density $\rho_p$ attain attain terminal velocities of
\begin{equation}
v_{t} \sim (g \rho_p s_p / \rho_g)^{1/2}
\end{equation}
in a background gas with density $\rho_g$. Since the distance over
which these particle attain their terminal speed, $L_{term} \sim
\rho_p s_p / \rho_g$, is smaller than the density scale height $\sim
c_{s, d}^2 / g$, particles larger than $\sim 1 mm$ cannot be carried
by the upwelling current on the day-side, and are left well beneath
the planet's photosphere. This potential channel for grain-gas
separation implies that it is very likely that the opacity in the
planet's atmosphere is well below that of the interstellar medium. A
detailed analysis of the grain evolution in the atmosphere of
short-period gas giants will be presented elsewhere.

We also presented a detailed comparison of our model to previous
dynamical models of \citet{showman2002}, \citet{cho2003},
\citet{burkert2005}, \citet{cooper2005}, \citet{cho2006}, and
\citet{langton2007}. The approaches among these groups varies
greatly, and we have attempted to highlight several of the crucial
differences. The most significant differences that we include are
solving the full fluid equations in all three-dimensions, rotation,
and our treatment of radiation via the flux-limited diffusion
method. Flux-limited radiation transport allowed for a self-consistent
treatment of the flow of radiation throughout the planet. This
treatment produced significantly shorter radiative timescales then
previously calculated, yielding much lower night-side temperatures.

The dynamics and heat distribution within the atmospheres giant
planets allow us to probe fundamental questions surrounding planet
formation. The diversity of planetary sizes must ultimately result
from interior structure variations arising from formation and
evolutionary processes. These atmospheres not only serve as a valuable
observable links to the interior, but may also play a role in
regulating planetary contraction rates. As suggested in
\citet{burrows2000} and \citet{burrows2006} an isothermal atmosphere
will result in decreased heat transfer and increased planetary
contraction timescales. However, a strong day-night temperature
difference may allow an avenue for cooling via the night-side. Based
on the simulations presented here, it is clear that the efficiency of
redistribution decreases with increasing opacity. High opacity
atmospheres, similar to interstellar values, will have cool night-side
temperatures, providing an avenue for internal heat loss and radial
contraction. If, as is expected, there is a significant amount of
grain growth and sedimentation, thus reducing the atmospheric opacity,
the planets night-side will also contain a large isothermal region,
possibly allowing for the retention of internal energy. Diversity in
atmospheric opacity may lead to diversity in planetary radii. With the
promise of improved observational techniques, including transit
spectroscopy and full phase light-curves, the relation between opacity
and temperature differential should be testable in the near future.

\acknowledgements We thank Geoff Bryden, Peter Bodenheimer, Katherine
Kretke, and Gordon Ogilvie for supplying the original version of our
numerical scheme, initial structural model, and many helpful
discussions. We would also like to acknowledge the use of NASA's High
End Computing Program computer systems. This work is partially
supported by NASA (NAGS5-11779, NNG04G-191G, NNG06-GH45G), JPL
(1270927), NSF(AST-0507424, PHY99-0794) and IGPP.

\bibliographystyle{aa}
\bibliography{ian}
\end{document}